\documentclass[aps,pra,amsmath,amssymb,tightenlines,epsfig,floatfix,twocolumn,superscriptaddress, longbibliography]{revtex4-1}

\usepackage{amsmath}
\usepackage{graphicx}
\usepackage{epstopdf}
\usepackage{dcolumn}	
\usepackage{bm}			
\usepackage{amsfonts}
\usepackage{xspace}
\usepackage{color}
\usepackage{multirow}
\usepackage{hyperref}
\usepackage{amssymb}
\usepackage{dutchcal}
\usepackage{amssymb}
\usepackage{mathtools}

\begin{document}
\newcommand{\psihat}{\ensuremath{\hat{\psi}}\xspace}
\newcommand{\psihatd}{\ensuremath{\hat{\psi}^{\dagger}}\xspace}
\newcommand{\ahat}{\ensuremath{\hat{a}}\xspace}
\newcommand{\Ham}{\ensuremath{\mathcal{H}}\xspace}
\newcommand{\ahatd}{\ensuremath{\hat{a}^{\dagger}}\xspace}
\newcommand{\bhat}{\ensuremath{\hat{b}}\xspace}
\newcommand{\bhatd}{\ensuremath{\hat{b}^{\dagger}}\xspace}
\newcommand{\boldr}{\ensuremath{\mathbf{r}}\xspace}
\newcommand{\dr}{\ensuremath{\,d^3\mathbf{r}}\xspace}
\newcommand{\tr}{\ensuremath{\,\mathrm{Tr}}\xspace}
\newcommand{\dk}{\ensuremath{\,d^3\mathbf{k}}\xspace}
\newcommand{\etal}{\emph{et al.\/}\xspace}
\newcommand{\ie}{i.e.\ }
\newcommand{\eq}[1]{Eq.\,(\ref{#1})\xspace}
\newcommand{\fig}[1]{Fig.\,\ref{#1}\xspace}
\newcommand{\abs}[1]{\left| #1 \right|}
\newcommand{\proj}[2]{\left| #1 \rangle\langle #2\right| \xspace}
\newcommand{\Qhat}{\ensuremath{\hat{Q}}\xspace}
\newcommand{\Qhatd}{\ensuremath{\hat{Q}^\dag}\xspace}
\newcommand{\phihatd}{\ensuremath{\hat{\phi}^{\dagger}}\xspace}
\newcommand{\phihat}{\ensuremath{\hat{\phi}}\xspace}
\newcommand{\boldk}{\ensuremath{\mathbf{k}}\xspace}
\newcommand{\boldp}{\ensuremath{\mathbf{p}}\xspace}
\newcommand{\boldsigma}{\ensuremath{\boldsymbol\sigma}\xspace}
\newcommand{\boldalpha}{\ensuremath{\boldsymbol\alpha}\xspace}
\newcommand{\grad}{\ensuremath{\boldsymbol\nabla}\xspace}
\newcommand{\parti}[2]{\frac{ \partial #1}{\partial #2} \xspace}
 \newcommand{\vs}[1]{\ensuremath{\boldsymbol{#1}}\xspace}
\renewcommand{\v}[1]{\ensuremath{\mathbf{#1}}\xspace}
\newcommand{\Psihat}{\ensuremath{\hat{\Psi}}\xspace}
\newcommand{\Psihatd}{\ensuremath{\hat{\Psi}^{\dagger}}\xspace}
\newcommand{\Vhatd}{\ensuremath{\hat{V}^{\dagger}}\xspace}
\newcommand{\Xhat}{\ensuremath{\hat{X}}\xspace}
\newcommand{\Xhatd}{\ensuremath{\hat{X}^{\dag}}\xspace}
\newcommand{\Yhat}{\ensuremath{\hat{Y}}\xspace}
\newcommand{\Jhat}{\ensuremath{\hat{J}}\xspace}
\newcommand{\Yhatd}{\ensuremath{\hat{Y}^{\dag}}\xspace}
\newcommand{\Uhat}{\ensuremath{\hat{U}^{\dag}}\xspace}
\newcommand{\jhat}{\ensuremath{\hat{J}}\xspace}
\newcommand{\lhat}{\ensuremath{\hat{L}}\xspace}
\newcommand{\Nhat}{\ensuremath{\hat{N}}\xspace}
\newcommand{\rhohat}{\ensuremath{\hat{\rho}}\xspace}
\newcommand{\ddt}{\ensuremath{\frac{d}{dt}}\xspace}
\newcommand{\nset}{\ensuremath{n_1, n_2,\dots, n_k}\xspace}
\newcommand{\Var}{\ensuremath{\mathrm{Var}}\xspace}
\newcommand{\Erf}{\ensuremath{\mathrm{Erf}}\xspace}
\newcommand{\sah}[1]{{\color{magenta}#1}}
\newcommand{\mk}[1]{{\color{red}#1}}


\title{Spin Squeezing of a Bose-Einstein Condensate via Quantum Non-Demolition Measurement for Quantum-Enhanced Atom Interferometry}
\author{Michail Kritsotakis}
\affiliation{Department of Physics and Astronomy, University of Sussex, Brighton BN1 9QH, United Kingdom}
\email{M.Kritsotakis@sussex.ac.uk}
\author{Jacob A.~Dunningham}
\affiliation{Department of Physics and Astronomy, University of Sussex, Brighton BN1 9QH, United Kingdom}
\author{Simon A.~Haine}
\affiliation{Department of Quantum Science, Research School of Physics and Engineering, The Australian National University, Canberra ACT 2601, Australia}

\begin{abstract}
We theoretically investigate the use of quantum non-demolition measurement to enhance the sensitivity of atom interferometry with Bose-condensed atoms. In particular, we are concerned with enhancing existing high-precision atom interferometry apparatuses, so restrict ourselves to dilute atomic samples, and the use of free-propagating light, or optical cavities in the weak-coupling regime.  We find the optimum parameter regime that balances between spin squeezing and atomic loss, and find that significant improvements in sensitivity are possible. Finally, we consider the use of squeezed light, and show that this can provide further boosts to sensitivity. 

\end{abstract}

\maketitle

\section{Introduction}
Atom interferometers are powerful tools for making precision measurements particularly in the realm of inertial navigation since they can provide sensitive measurements of accelerations and rotations with very low baseline drift \cite{Cronin:2009, Robins:2013}.  A lot of interest has therefore developed in finding ways of improving their performance to gain advantage in different applications. It has been shown that Bose-condensed atomic sources can outperform thermal sources due to their narrow momentum linewidth, despite their reduced atomic flux \cite{Johnsson:2007b, Debs:2012, Szigeti:2012, Mcdonald:2013, Kritsotakis:2018}.  The use of non-classical atomic states such as spin-squeezed states can offset this reduction in flux even further by allowing for sensitivities beyond the shot-noise limit (SNL) \cite{Wineland:1992, Kitagawa:1993, Sorensen:2001, Pezze_review:2018}. In this paper we investigate the use of quantum non-demolition (QND) measurements in collections of Bose-condensed atoms to generate quantum states that could be used to enhance their precision in a range of metrology schemes. The mechanism for generating quantum enhanced many-atom states can be broadly classified into two categories: Those that use atom-atom interactions \cite{Kitagawa:1993, Duan:2000, Pu:2000, Sorensen:2002, Micheli:2003, Kheruntsyan:2005b, Johnsson:2007a, Li:2009, Mirkhalaf:2018}, and those that use atom-light interactions \cite{Kuzmich:1997, Kuzmich:1998, Moore:1999, Kuzmich:2000, Jing:2000, Fleischhauer:2002b, Haine:2005, Haine:2005b, Echaniz:2005, Haine:2006b, Hammerer:2010, Haine:2013, Puentes:2013, Szigeti:2014b, Tonekaboni:2015, Haine:2015, Haine:2015b, Haine:2016, Salvi:2018}. While several experiments have demonstrated non-classical states generated through atom-atom interactions \cite{Esteve:2008, Gross:2010, Riedel:2010, Lucke:2011, Hamley:2012, Strobel:2014, Muessel:2014, Kruse:2016, Linnemann:2016, Zou:2018}, so far these have been restricted to small numbers of atoms, and have not been applied to atom interferometry capable of inertial measurements. This is partly because the atom-atom interactions required for the generation of the entanglement create unavoidable multimode-dynamics which inhibit mode-matching, \cite{Li:2009, Haine:2009, Haine:2011, Opanchuk:2012, Haine:2014}, and phase diffusion \cite{Nolan:2016, Haine:2018}. Quantum entanglement through atom-light interactions, which are free to operate in regimes where the effects of atom-atom interactions are negligible, have also been successfully demonstrated. In particular, the use of light to perform QND measurements of the collective atomic spin has shown significant spin-squeezing \cite{Appel:2009, Louchet-Chauvet:2010, Schleier-Smith:2010, Schleier-Smith:2010b, Leroux:2010, Koshchorreck:2010, Sewell:2012, Sewell:2014, Hosten:2016}.  So far, these experimental demonstrations have been restricted to cold thermal atoms. In this work, we focus on Bose-condensed sources, with the motivation of implementing this quantum enhancement technique on existing high-precision, large space-time area atomic gravimetry set-ups, such as \cite{Altin:2013}. In particular, the requirement that the Bose-Einstein condensate (BEC) is expanded before the atomic beam-splitting process dictates a minimum spatial size of the source, and prevents excessively elongated samples such as in \cite{Sewell:2012}. Furthermore, we restrict ourselves to freely propagating light, and find the optimum parameter regime which balances the spin-squeezing and atomic loss caused by spontaneous emission. We also consider the use of optical cavities, but restrict ourselves to cavities that are assembled outside the vacuum chamber, so are inherently low-finesse with weak atom-light coupling due to the large cavity volume. We also consider the use of squeezed light to further enhance the sensitivity.

This paper is structured as follows. In section \ref{sec2} we review atom interferometry and quantify how spin-squeezing via QND measurements improves the sensitivity. In section \ref{single_mode_light_sec} we introduce a simple model of QND squeezing which allows us to make some simple analytic scaling predictions. In section \ref{scheme_sec} we present our full model including a freely-propagating multimode optical field and decoherence due to spontaneous emission. In section \ref{sec5} we derive approximate analytic solutions to this model, and in section \ref{numerical} we analyse the system numerically. In section \ref{sec7} we investigate how the use of squeezed light affects the behaviour. In section \ref{sec8} we investigate the use of an optical cavity. 

\section{Using QND measurements to enhance the sensitivity of a Mach-Zehnder Interferometer. }\label{sec2}
Atom interferometers used to measure accelerations and rotations are usually based on the Mach-Zehnder (MZ) configuration \cite{Kasevich:1992, Riehle:1991}. Starting with an ensemble of atoms with two stable ground states, labeled $|1\rangle$ and $|2\rangle$, a $\frac{\pi}{2}$ pulse, or `beamsplitter', implemented by a two-photon Raman transition is used to place each atom in an equal superposition of these states, while transferring momentum $\hbar \boldk_0$ to the state $|2\rangle$ component, where $\boldk_0$ is determined by the difference in wavevectors of the two Raman lasers.  The atoms then evolve for a period of time $T$, before a second Raman transition implements a $\pi$ pulse, or `mirror'. After a second period of time $T$, a second $\frac{\pi}{2}$ beamsplitter pulse is implemented, and the number difference is read-out. Such a system is conveniently described by introducing the pseudo-spin operators $\hat{J}_x$, $\hat{J}_y$, $\hat{J}_z$, defined by
\begin{equation}
\hat{J}_j = \frac{1}{2}\int \hat{\boldsymbol{\psi}}^\dag(\boldr)\sigma_j\hat{\boldsymbol{\psi}}(\boldr) \, \dr
\end{equation} 
where $\sigma_j$ is the $j$th Pauli matrix, and 
\begin{equation}
\hat{\boldsymbol{\psi}}(\boldr) = \begin{bmatrix} \hat{\psi}_1(\boldr) \\ e^{-i \boldk_0\cdot \boldr }\hat{\psi}_2(\boldr) \end{bmatrix} \, ,
\end{equation}
where $\hat{\psi}_{n}(\boldr)$ are the bosonic field operators that annihilate a particle at point $\boldr$ from state $|n\rangle$, satisfying the usual commutation relations
\begin{subequations}
\begin{align}
\left[\hat{\psi}_n(\boldr) \, , \, \hat{\psi}_m^\dag(\boldr^\prime)\right] &=  \delta_{n,m}\delta(\boldr-\boldr^\prime) \\
\left[\hat{\psi}_n(\boldr) \, , \, \hat{\psi}_m(\boldr^\prime)\right] &=  \left[\hat{\psi}^\dag_n(\boldr) \, , \, \hat{\psi}^\dag_m(\boldr^\prime)\right] =0 \,.
\end{align}
\end{subequations}
These operators obey the SU(2) commutation relations:
\begin{equation}
\left[\hat{J}_x, \hat{J}_y\right] = i\hat{J}_z, \, \left[\hat{J}_y, \hat{J}_z\right] = i\hat{J}_x, \, \left[\hat{J}_z, \hat{J}_x\right] = i\hat{J}_y \, .
\end{equation}
It can be shown that the MZ interferometer described above performs the operation 
\begin{equation}
\jhat_k = e^{i \jhat_y \theta} \jhat_k(0) e^{-i \jhat_y\theta} \, .
\end{equation}
where $\theta$ is the phase difference that has accumulated between the two arms of the interferometer, and $\hat{J}_k(0)$ are the operators before the pulse sequence \cite{Schleich:2013, Kleinert:2015, Kritsotakis:2018, Bertoldi:2019}. For a gravimeter, $\theta = \boldsymbol{g}\cdot \boldk_0 T^2$ where $\boldsymbol{g}$ is the gravitational field. The task of estimating $g$, the magnitude of the gravitational field parallel to $\boldk_0$ then comes down to our ability to estimate $\theta$. That is, $\Delta g = \Delta \theta/ |\boldk_0| T^2$.    

For a particular measurement signal $\hat{S}$, the sensitivity is given by
\begin{equation}
\Delta \theta = \sqrt{\frac{\Var(\hat{S})}{(\partial_\theta \langle \hat{S}\rangle)^2}} \, .
\end{equation}
Choosing $\hat{S}_1 = \jhat_z$ for $\hat{S}$, the number difference at the output of the interferometer, we find
\begin{equation}
\hat{J}_z =  \hat{J}_z(0)\cos(\theta) - \hat{J}_x(0)\sin(\theta) \, .
\end{equation}
Operating around $\theta\approx 0$, we find
\begin{equation}
\Delta \theta =  \sqrt{\frac{\Var{(\jhat_z)}}{(\partial_\theta \langle \jhat_x\rangle)^2}} \, .
\end{equation}
Choosing an initial state as $N_a$ uncorrelated atoms in an equal superposition of $|1\rangle$ and $|2\rangle$, ie, a coherent spin state \cite{Radcliffe:1971},
\begin{equation}
|\Psi\rangle = \left(\frac{1}{\sqrt{2}}\left(|1\rangle + |2\rangle \right)\right)^{\otimes N_a} \, ,
\end{equation}
 we find
\begin{equation}
\Delta \theta =  \frac{1}{\sqrt{N_a}} \, ,\label{SNL}
\end{equation}
which is the shot-noise limit (SNL). This is the best possible sensitivity for any uncorrelated state. That is, any state of the form $|\Psi\rangle= (c_1 |1\rangle + c_2|2\rangle)^{\otimes N_a}$. Equation \ref{SNL} motivates the introduction of the spin-squeezing parameter $\xi_s$, defined by
\begin{align}
\xi_s = \sqrt{N_a}\frac{\sqrt{\mathrm{Var}(\hat{S})}}{|\partial_\theta\langle\hat{S}\rangle|}
\label{xi_s}
\end{align}
such that
\begin{align}
\Delta\theta = \frac{\xi_s}{\sqrt{N_a}} \, .
\label{Delta_phi}
\end{align}

The use of input states with quantum correlations such that $\xi_s <1$ gives sensitivities better than the SNL. This can be done by creating atom-atom entanglement, but can also be done by creating entanglement between the atoms and some auxiliary field, such as an optical beam. By measuring both fields together, it is possible to create a signal with reduced fluctuations and therefore increased sensitivity. Specifically, by measuring the combined signal
\begin{align}
\hat{S}_2 = \hat{J}_z - \hat{J}_z^{\mathrm{inf}}
\end{align}
where $\hat{J}_z^{\mathrm{inf}}  = G\hat{S}_b$ represents an inference of the population difference, based on measurements of some optical observable $\hat{S}_b$. The constant $G$  is a proportionality factor, which is found by minimizing the variance of the total signal $\mathrm{Var}(\hat{S}_2)$ with respect to $G$:
\begin{align}
G=\frac{\mathrm{Cov}\left(\hat{J}_z,\hat{S}_b\right)}{\mathrm{Var}(\hat{S}_b)}
\end{align}
which gives:

\begin{align}
\mathrm{Var}(\hat{S}_2) = \mathrm{Var}(\hat{J}_z) - \frac{\mathrm{Cov}^2(\hat{J_z},\hat{S}_b)}{\mathrm{Var}(\hat{S}_b)}
\end{align}

Hence, creating atom-light entanglement and measuring the appropriate light observable in such a way that $\frac{\mathrm{Cov}^2(\hat{J_z},\hat{S}_b)}{\mathrm{Var}(\hat{S}_b)} > 0$, yields a reduced signal variance $\mathrm{Var}(\hat{S}_2) < \Var(\hat{J}_z) = \Var(\hat{S}_1)$, increasing the sensitivity over purely measuring the population difference between the two interferometer modes. As the optical observables are unaffected by the MZ sequence, which only acts on atomic degrees of freedom, at $\theta=0$ we have $\Delta \theta = \xi_{s_2}/\sqrt{N_a}$, where

\begin{align}
\xi_{s_2} = \frac{\sqrt{N_a}\sqrt{\mathrm{Var}(\hat{S}_2)}}{|\langle\hat{J}_x\rangle|} \, . 
\label{signal_2}
\end{align}
If the Hamiltonian responsible for the atom-light entanglement commutes with $\jhat_z$, then this is an example of a QND measurement, as there is no measurement back-action on the observable being measured. In the next section we model the atom-light interaction, and quantify how the appropriate choice of $\hat{S}_b$ improves the sensitivity.

\section{Simple Model: Single mode Light fields}\label{single_mode_light_sec}
\begin{figure}[h]
	\begin{center}
		\includegraphics[width=0.5\columnwidth]{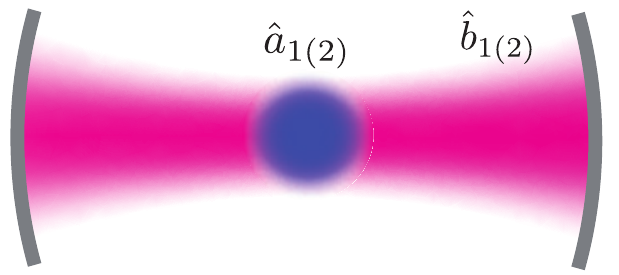}
		\caption{Simplified scheme showing QND entanglement via an atom-light interaction. An optical mode represented by annihilation operator $\bhat_{1(2)}$ interacts with an ensemble of Bose-condensed atoms (annihilation operator $\ahat_{1(2)}$).}
		\label{simple_scheme}
	\end{center}
\end{figure}

In order to demonstrate how QND squeezing affects the sensitivity, we begin with a simplified model where we make the single mode approximation for both the atomic fields and optical fields. Assuming an ensemble of two-level atoms in the ground motional state of a trapped BEC, with each level interacting with a far-detuned laser beam, as described in figure \ref{simple_scheme}. 

The simplified Hamiltonian for the system is 
\begin{equation}
\hat{H}_{int} = - \hbar\chi_{\mathrm{sm}}(\hat{a}_1^\dagger\hat{a}_1\hat{b}_1^\dagger\hat{b}_1 + \hat{a}_2^\dagger\hat{a}_2\hat{b}_2^\dagger\hat{b}_2) \, ,
\end{equation}
where $\chi_{\mathrm{sm}}$ indicates the interaction strength between the atoms and the light in our simple model. Also, $\ahat_{j} = \int u^*_{0}(\boldr) \psihat_j(\boldr) \dr$ annihilates an atom from the ground motional state of the BEC (spatial wavefunction $u_0(\boldr)$), and $\bhat_j$ annihilates a photon from the optical mode interacting with atomic state $|j\rangle$.  The atomic and light operators satisfy $[\ahat_i,\ahatd_j] = \delta_{ij}$ and $[\bhat_i, \bhatd_j] = \delta_{ij}$ respectively. As both $\ahatd_j\ahat_j$ and $\bhatd_j\bhat_j$ commute with the Hamiltonian, the solution to the Heisenberg equations of motion for the system are
\begin{subequations}
\begin{align}
\hat{a}_j(t) &= \hat{a}_j(0)e^{i\chi_{\mathrm{sm}}\hat{b}_j^\dagger(0)\hat{b}_j(0) t} \label{asol1} \\
\hat{b}_j(t) &= \hat{b}_j(0)e^{i\chi_{\mathrm{sm}}\hat{a}_j^\dagger(0)\hat{a}_j(0) t} \label{bsol1}
\end{align}
\end{subequations}
Examining the form of \eq{bsol1}, we see that the phase of the optical mode is correlated with the population of the corresponding atomic mode. This motivates us to examine $\hat{Y}_{b_j}$, where $\hat{Y}_{b_j} = i\left(\hat{b}_j - \hat{b}_j^\dagger\right)$ is the phase quadrature of the light field. After making the small angle approximation $\chi_{\mathrm{sm}} t \hat{a}_j^\dagger\hat{a}_j << 1$ we find 
\begin{align}
\hat{Y}_j(t) \approx\hat{Y}_{j0} - \chi_{\mathrm{sm}}\hat{a}_j^\dagger\hat{a}_jt\hat{X}_{j0}
\label{phase_quad_simple_model}
\end{align}
where $\hat{Y}_{j0} = i(\hat{b}_j(0) -\hat{b}_j^\dagger(0))$ and $\hat{X}_{j0} = \hat{b}_j(0) + \hat{b}_j^\dagger(0)$, and notice that $\hat{Y}_j(t) \propto \hat{N}_{a_j}$. Hence we can make an inference about the atomic population difference by measuring the difference of the two phase quadratures. Setting $\hat{S}_b = \hat{Y}_2 - \hat{Y}_1$, and using a Glauber coherent state $|\beta_j\rangle$, with $\mathrm{Im}(\beta)=0$ as the initial state for our optical modes, we find
\begin{align}
\Var(\hat{S}_b(t)) \approx 2 + 4\chi_{\mathrm{sm}}^2N_{\mathrm{ph}}N_at^2 \, ,
\end{align} 
and
\begin{align}
\Var(\hat{S}_2(t)) = \frac{N_a}{4}\left(1 - \frac{2\chi_{\mathrm{sm}}^2N_aN_{\mathrm{ph}}t^2}{1 + 2\chi_{\mathrm{sm}}^2N_aN_{\mathrm{ph}}t^2}\right)
\end{align}
where $N_{\mathrm{ph}} = |\beta_1|^2 +|\beta_2|^2$ is the expectation value of the number of photons. Using this in \eq{signal_2} we find
\begin{align}
\xi_{s_2} = e^{\chi_{\mathrm{sm}}^2N_{\mathrm{ph}}t^2}\left(1 - \frac{2\chi_{\mathrm{sm}}^2N_aN_{\mathrm{ph}}t^2}{1 + 2\chi_{\mathrm{sm}}^2N_aN_{\mathrm{ph}}t^2}\right)^{1/2} \, .
\end{align}

\begin{figure}[h]
	\begin{center}
		\includegraphics[width=1.0\columnwidth]{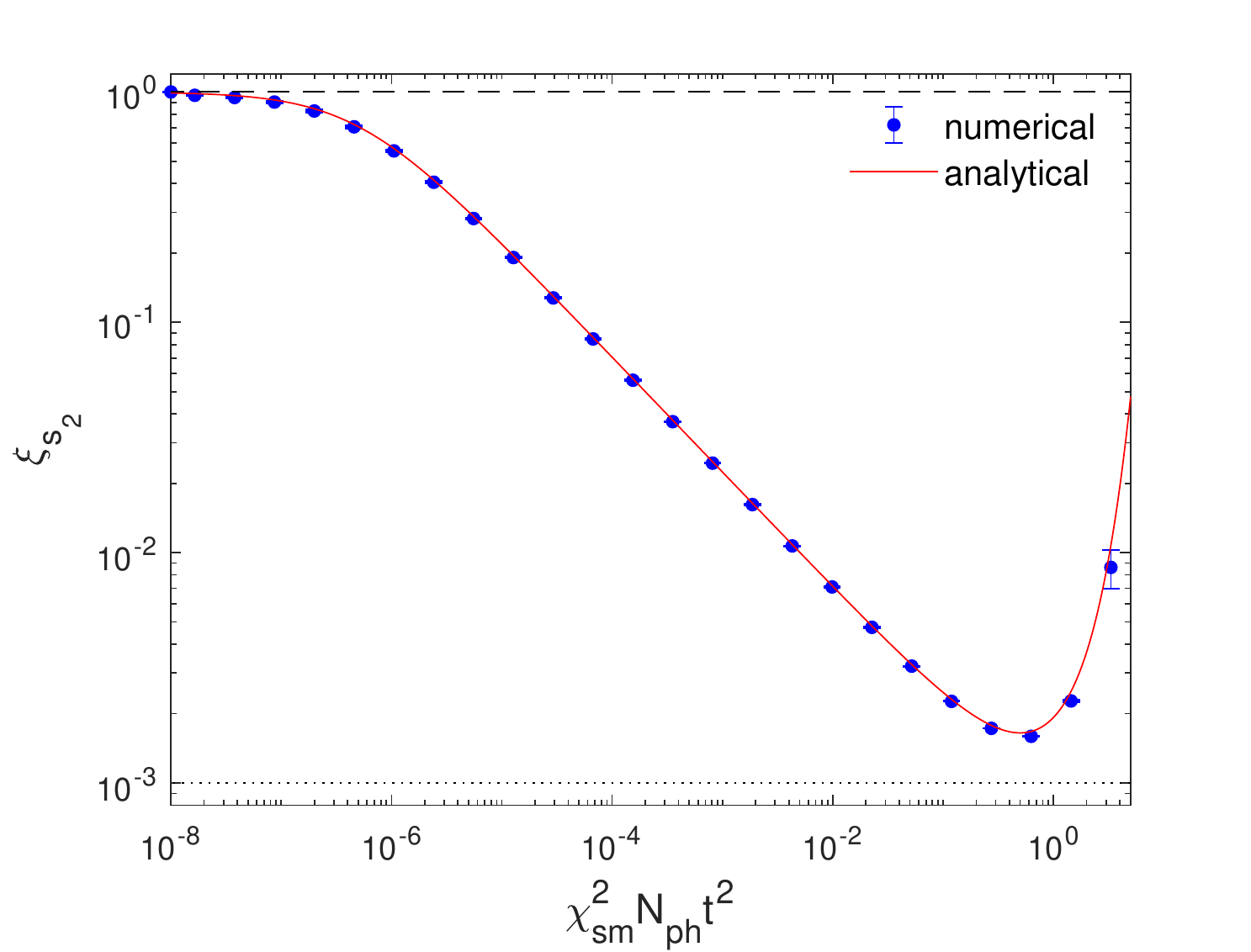}
		\caption{Simple model: Analytical (red line) and numerical calculation (blue dots) of $\xi_{s_2}$ with respect to the collective parameter $\chi_{\mathrm{sm}}^2N_{\mathrm{ph}}t^2$. The black dashed line and black dotted line represent the SNL and the Heisenberg limit respectively. The error bars were calculated by taking the standard deviation over many different iterations of the system dynamics.}
		\label{super_simple_model}
	\end{center}
\end{figure}

We notice in Fig.~[\ref{super_simple_model}] that we obtain better sensitivities for our signal compared to the SNL, indicating that we have created a spin squeezed state. We find the optimum value for the number of photons $N_{\mathrm{ph}}^{\mathrm{opt}} = \frac{1}{2\chi_{\mathrm{sm}}^2t^2}$ which gives the minimum value $\left(\xi_{s_2}^{\mathrm{sm}}\right)_{\mathrm{min}} = \sqrt{\frac{e}{N_a}}$. 

This section demonstrates that this kind of atom-light interaction creates an atomic spin squeezed state and consequently boosts the interferometer's performance. In the following section we model the system more rigorously, using the freely propagating light field and including the effects of atomic spontaneous emission. 



\section{Detailed model describing atom-light interaction}\label{scheme_sec}

\begin{figure}[h]
	\begin{center}
		\includegraphics[width=0.7\columnwidth]{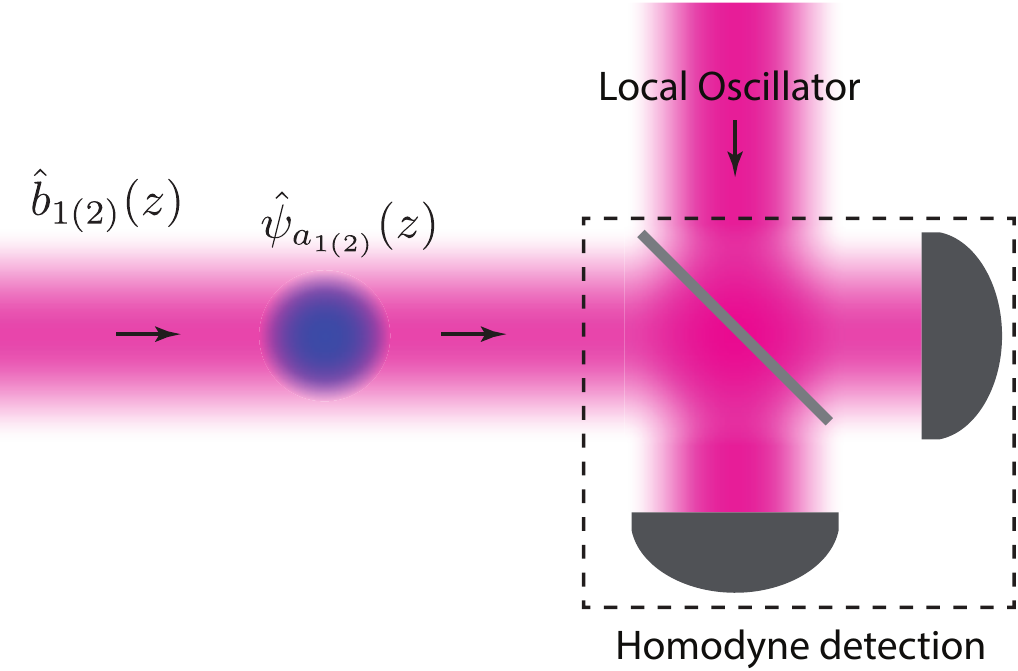}
		\caption{Schematic of the free-space QND scheme. After interacting with the atomic ensemble, the freely propagating optical field is measured via homodyne detection.}
		\label{free_space_scheme}
	\end{center}
\end{figure}

We now consider a more detailed model that more accurately captures the relevant physics. In particular, in order to model propagating laser beams, we require a multi-mode model for the optical fields (see figure \ref{free_space_scheme}). We also include spontaneous emission from the excited atomic states, which will limit the amount of QND squeezing in practice. 

\subsection{Equations of motion describing atom-light interaction}
\begin{figure}[h]
	\begin{center}
		\includegraphics[width=0.7\columnwidth]{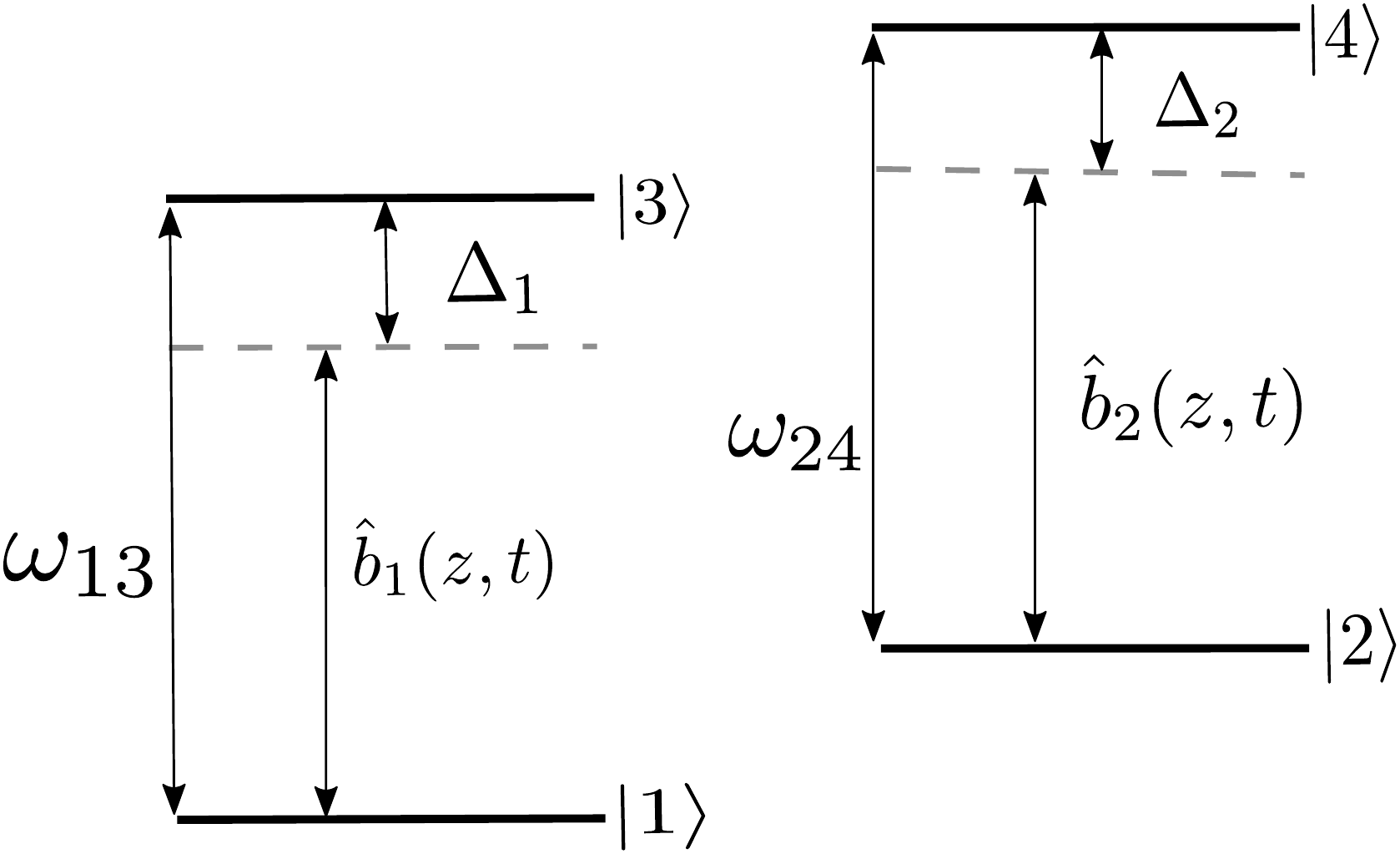}
		\caption{Atomic energy diagram of the two 2-level systems. Each atom is placed in a superposition of electronic states $|1\rangle$ and $|2\rangle$, with excited states $|3\rangle$ and $|4\rangle$. Two independent lasers (annihilation operator $\bhat_1$ and $\bhat_2$) are detuned from the $|1\rangle \rightarrow |3\rangle$ and $|2\rangle \rightarrow |4\rangle$ transitions by detuning $\Delta_1$ and $\Delta_2$, respectively.}
		\label{atomic_energy_levels}
	\end{center}
\end{figure}

We assume an ensemble of Bose-condensed atoms with two electronic states $|1\rangle$ and $|2\rangle$, coupled to excited states $|3\rangle$ and $|4 \rangle$ respectively (Fig.~\ref{atomic_energy_levels}). The coupling is achieved by far-detuned lasers, which are described by annihilation operators $\hat{b}_1(z,t)$ and $\hat{b}_2(z,t)$, satisfying the commutation relations $[\hat{b}_i(z,t),\hat{b}_j^\dagger(z',t)] = \delta_{ij}\delta(z-z')$ for $i, j=1,2$. We assume both optical fields have narrow linewidths compared to the natural linewidths of the atomic transitions, with central frequencies given by $\omega_{L_1} = \omega_{13} -\Delta_1$ and $\omega_{L_2} = \omega_{24} -\Delta_2$, where $\Delta_{1}$ and $\Delta_{2}$ are the detunings from the $|1\rangle \rightarrow |3\rangle$ and $|2\rangle \rightarrow |4\rangle$ transitions, respectively.  The Hamiltonian for the total system after making the rotational-wave-approximation (RWA) is
\begin{align}
\hat{H}_{\mathrm{tot}} &= \hbar\int_{-\infty}^{\infty}dz\left(\omega_{13}\hat{\psi}_3^{\dagger}(z,t)\hat{\psi}_3(z,t) + \omega_{24}\hat{\psi}_4^{\dagger}(z,t)\hat{\psi}_4(z,t)\right) \nonumber\\ &- i\hbar c\int_{-\infty}^{\infty}\hat{b}_1^{\dagger}\partial_z\hat{b}_1dz - i\hbar c\int_{-\infty}^{\infty}\hat{b}_2^{\dagger}\partial_z\hat{b}_2dz  \nonumber\\
& + \; \hbar g_{13}\int_{-\infty}^{\infty}\left(\hat{\psi}_1^{\dagger}(z,t)\hat{\psi}_3(z,t)\hat{b}_1^\dagger(z,t) + \mathrm{h.c}\right)dz \nonumber\\ 
& + \; \hbar g_{24}\int_{-\infty}^{\infty}\left(\hat{\psi}_2^{\dagger}(z,t)\hat{\psi}_4(z,t)\hat{b}_2^\dagger(z,t) + \mathrm{h.c}\right)dz \, ,
\label{Hamilt}
\end{align}
where $\hat{\psi}_i(z,t)$ is the field operator which annihilates an atom from atomic state $|i\rangle$ at position $z$, and $g_{13}=\frac{d_{13}}{\hbar}\left(\frac{\hbar \omega_{L_1}}{2\epsilon_0 A}\right)^{1/2}$ and $g_{24}=\frac{d_{24}}{\hbar}\left(\frac{\hbar \omega_{L_2}}{2\epsilon_0 A}\right)^{1/2}$ are the atom-light coupling constant, where $d_{13}=-e\langle 3|\hat{\boldr} |1\rangle$ and  $d_{24}=-e\langle 4|\hat{\boldr} |2\rangle$ are the dipole moment matrix elements for the atomic transitions $|1\rangle\rightarrow|3\rangle$ and $|2\rangle\rightarrow|4\rangle$ respectively, $A$ is the transverse quantization area of the light beam and $c$ is the speed of light.

For simplicity in the following we will present the Heisenberg equations of motion just for one two-level system \{$|1\rangle\rightarrow|3\rangle, \; \hat{b}_1(z,t)$\}, since the two systems are de-coupled in the sense that the Heisenberg equations of motion for $|1\rangle\rightarrow|3\rangle$ and $|2\rangle\rightarrow|4\rangle$ are independent. The corresponding equations hold for the second two-level system \{$|2\rangle\rightarrow|4\rangle, \; \hat{b}_2(z,t)$\} as well. 


We incorporate spontaneous emission as a Langevin term in our Heisenberg equation of motion, by coupling the atoms being in their excited state to a reservoir of vacuum electromagnetic modes, which is then traced over, described by the Hamiltonian $\hat{H}_{\mathrm{bath}}=\hbar\int_{-\infty}^{\infty}dz\int_{-\infty}^{\infty}d\omega \, \omega \, \hat{d}^\dagger(\omega,z)\hat{d}(\omega,z)$, where $\hat{d}(\omega,z)$ is the continuous in space and frequency annihilation operator of the bath satisfying $[\hat{d}(\omega,z),\hat{d}^\dagger(\omega',z')] = \delta(\omega -\omega')\delta(z-z')$. Hence, the equation of motion for $\hat{\psi}_3(z,t)$ in the presence of this Langevin term \cite{Langevin_Gardiner} is 
\begin{align}
\partial_t\hat{\psi}_3(z,t)=-\frac{i}{\hbar}&\left[\hat{\psi}_3(z,t),\hat{H}_{\mathrm{tot}}\right] \, + \nonumber \\
-\, &\left(\frac{\gamma_3}{2}\hat{\psi}_3(z,t) + \sqrt{\gamma_3}\hat{d}_{1_\mathrm{in}}(z,t)\right) \, ,
\label{Lang_eq_psi3}
\end{align}
where $\gamma_3$ is the spontaneous emission rate from the excited state and $\hat{d}_{1_\mathrm{in}}(z,t)=\frac{1}{\sqrt{2\pi}}\int_{-\infty}^{\infty}d\omega e^{-i\omega(t-t_0)}\hat{d}_0(\omega,z)$ is the standard Langevin noise term depending on the value of the bath operator at the initial time point $t_0$, $\hat{d}(\omega,z,t=t_0)=\hat{d}_0(\omega,z)$.  After moving to a rotating reference frame, with respect to the central frequency of the light field, $\omega_{L_1}$, we adiabatically eliminate the excited state field operator $\psihat_3$, \cite{adiabatic_elimination_Molmer}. Thus, the Heisenberg equations of motion for $\hat{\psi}_1(z,t)$ and $\hat{b}_1(z,t)$ are
\begin{subequations}
\begin{align}
 \partial_t\hat{\psi}_1(z,t) &=ig_{13}^2\frac{\Delta_1+i\frac{\gamma_3}{2}}{\Delta_1^2+\frac{\gamma_3^2}{4}}\hat{b}_1^\dagger(z,t)\hat{b}_1(z,t)\hat{\psi}_1(z,t) \nonumber \\ 
 &+ g_{13}\frac{\sqrt{\gamma_3}}{\Delta_1-i\frac{\gamma_3}{2}}\hat{b}_1^\dagger(z,t)\hat{d}_{1_\mathrm{in}}(z,t) ,
  \label{psi1_dif_eq_se} \\
 \left(\frac{1}{c}\partial_t+\partial_z\right)\hat{b}_1(z,t) &= i\frac{g_{13}^2}{c}\frac{\Delta_1+i\frac{\gamma_3}{2}}{\Delta_1^2+\frac{\gamma_3^2}{4}}\hat{\psi}_1^\dagger(z,t)\hat{\psi}_1(z,t)\hat{b}_1(z,t) \nonumber\\[5pt] &+ \frac{g_{13}}{c}\frac{\sqrt{\gamma_3}}{\Delta_1-i\frac{\gamma_3}{2}}\hat{\psi}_1^\dagger(z,t)\hat{d}_{1_\mathrm{in}}(z,t) .
 \label{b1_dif_eq_se}
 \end{align}
\end{subequations}

We solve the equation for the light field by making the substitution $z \rightarrow z + ct$. As the timescale for the atomic dynamics is much slower than the timescale for the light to cross the atomic sample, we make the approximation that the light moves between two arbitrary points $z_B$ to $z_C$ instantaneously, i.e $\hat{b}^\dag(z_B,t)\hat{b}(z_B,t) = \hat{b}^\dag(z_C,t)\hat{b}(z_C,t)$, as long as there is no atom-light interaction in $[z_B,z_C]$.  In addition, as our system is a Bose-Einstein condensate, we assume that all the atoms are in the ground motional state of the trap, which allows us to make the single mode approximation $\hat{\psi}_1(z,t)=u_0(z)\hat{a}_1(t)$. Assuming $\int_{z_L}^{z_R}|u_0(z)|^2 dz \approx 1$ for points $z_L$ and $z_R$ sufficiently far to the left and right of the atomic sample respectively, we can write
\begin{align}
\hat{b}_1(z_R,t) = \hat{b}_{01}(t)e^{i\frac{g_{13}^2}{c}(\Omega + i\Gamma)\hat{a}_1^\dagger(t)\hat{a}_1(t)} \; + \nonumber \\
+ \; \frac{g_{13}}{c}\frac{\sqrt{\gamma_3}}{\Delta_1 - i\gamma_3/2}\hat{a}_1^\dagger(t)\hat{q}_{1_{\mathrm{in}}}(t) \label{sol_b1}
\end{align}

where we have considered the same motional function for the Langevin noise $\hat{d}_{1_\mathrm{in}}(z,t) = u_0(z)\hat{q}_{1_\mathrm{in}}(t)$. We have also defined $\hat{b}_{01}(t) = \hat{b}_1(z_L,t)$, and $\Omega\equiv\frac{\Delta_1}{\Delta_1^2 + \gamma_3^2/4}$, $\Gamma_3 \equiv \frac{\gamma_3/2}{\Delta_1^2 + \gamma_3^2/4}$ for notation simplicity. 
%
%
%
%
%
%
%
%
%
%
%
In order to find a simpler form for the atomic equation, Eq.~(\ref{psi1_dif_eq_se}), we make the approximation that $\hat{b}^\dag_1(z,t)\hat{b}_1(z,t) \approx \hat{b}^\dag_1(z_L,t)\hat{b}_1(z_L,t)$, i.e. the number of photons in the mode does not change to a good approximation.
%
%
Hence, after making the single mode approximation again we obtain:
\begin{align}
\partial_t\hat{a}_1(t) &=ig_{13}^2\left(\Omega + i\Gamma\right)\hat{b}_{01}^\dagger(t)\hat{b}_{01}(t)\hat{a}_1(t) \nonumber \\ 
&+ g_{13}\frac{\sqrt{\gamma_3}}{\Delta_1-i\frac{\gamma_3}{2}}\hat{b}_{01}^\dagger(t)\hat{q}_{1_\mathrm{in}}(t) \, .
\end{align}

\subsection{Measurement of the Optical Observables}\label{measurement_sec}
As in section \ref{single_mode_light_sec}, we notice that Eq.~(\ref{sol_b1}) indicates correlations between the atomic number and the phase of the light. We can define the phase quadrature for our multimode light field by selecting one specific mode. Specifically, we define $\hat{Y}_{\mathcal{b}_1} = i\left(\hat{\mathcal{b}}_1 - \hat{\mathcal{b}}_1^\dagger\right)$ where
\begin{align}
\hat{\mathcal{b}}_1 = \int_{0}^{\tau} u_{\mathrm{LO}}^*(t)\hat{b}_1(z_D,t)dt
\end{align}
where $z_D$ is the position of the photo-detector. Also, $u_{\mathrm{LO}}(t)$ corresponds to the temporal mode shape of the local oscillator used in the homodyne detection \cite{Bachor:2004}, satisfying 
\begin{align}
\int_{0}^{\tau}|u_{\mathrm{LO}}(t)|^2dt = c 
\end{align}
which ensures $[\hat{\mathcal{b}}_1,\hat{\mathcal{b}}_2^\dagger] = 1$ and consequently $[\hat{X}_{\mathcal{b}_1},\hat{Y}_{\mathcal{b}_1}] = -2i$, where $\hat{X}_{\mathcal{b}_1} = \hat{\mathcal{b}}_1 + \hat{\mathcal{b}}_1^\dagger$ is the corresponding amplitude quadrature of $\hat{\mathcal{b}}_1$. The most appropriate choice of local oscillator for this scheme is one with constant intensity with the frequency matched to the carrier frequency of our optical field, i.e. 
\begin{equation}
u_{\mathrm{LO}}(t) = \sqrt{\frac{c}{\tau}}.
\end{equation}


%
%
%
%

\section{Approximate Analytic Solutions}\label{sec5}
We can obtain an analytical estimate of the quantum-enhancement parameter, $\xi_{s_2}$, after  making some approximations. Here we briefly present the basic intermediate steps we made in order to find out $\xi_{s_2}$, with and without spontaneous emission. A much more detailed presentation of these calculations can be found in the Appendices \ref{appendix_intro} - \ref{appendix_final}. For simplicity we assume that the atom-light interaction strengths as well as the detunings are the same for the two atomic transitions, i.e $g_{13} = g_{24} = g$ and $\Delta_1 =\Delta_2 = \Delta$ respectively. We also consider that initially the atoms and the light fields are in coherent states with the same amplitudes for the two atomic levels $\hat{a}_{1(2)}(0)|\alpha_{1(2)}\rangle = \sqrt{\frac{N_a}{2}}|\alpha_{1(2)}\rangle$ and for the light $\hat{b}_{01}(t)|\beta_1\rangle = \beta_0|\beta_1\rangle$, $\hat{b}_{02}(t)|\beta_2\rangle = \beta_0|\beta_2\rangle$ where we also assume that $\beta_0 = \beta_0^*$.

\subsection{No Spontaneous emission}
Ignoring the effect of spontaneous emission (ie, $\gamma_3=0$) vastly simplifies the problem and allows easy comparison with the simple single-mode model of section \ref{single_mode_light_sec}. In this case, the calculation of the atomic expectation values we are interested in is quite straightforward:
\begin{align}
\langle\hat{N}_{a_1}(t)\rangle = \frac{N_a}{2},
\qquad
\langle\hat{N}_{a_1}^2(t)\rangle = \frac{N_a}{2}\left(1 + \frac{N_a}{2}\right)
\label{atom_exp_np}
\end{align}
We can also find the phase quadrature operator by making the small angle approximation $\frac{g^2}{c\Delta}\hat{a}_1^\dagger(t)\hat{a}_1(t) << 1$:

\begin{align}
\hat{Y}_1(\tau) \approx \hat{Y}_{1_{\mathrm{in}}}(\tau) - \frac{g^2}{\sqrt{c\tau}\Delta}\hat{a}_1^\dagger(\tau)\hat{a}_1(\tau)\int_{0}^{\tau}\left(\hat{b}_{01}(t) + \hat{b}_{01}^\dagger(t)\right)dt
\label{phase_op_ns}
\end{align}

where $\hat{Y}_{1_{\mathrm{in}}}(\tau) = i\frac{\sqrt{c}}{\sqrt{\tau}}\int_{0}^{\tau}\left(\hat{b}_{01}(t) - \hat{b}_{01}^\dagger(t)\right)dt$.

Here we clearly notice that $\hat{Y}_1 \propto \hat{N}_{a_1}$. That supports our choice for the light signal to be $\hat{J}_z^{\mathrm{inf}} \propto \hat{S}_b = \hat{Y}_2 - \hat{Y}_1$. Now using Eq.~({\ref{atom_exp_np}) and (\ref{phase_op_ns}) we can calculate:

\begin{align}
&\mathrm{Var}(\hat{S}_b) \approx 2\mathrm{Var}(\hat{Y}_1(\tau)) \approx 2 + 4\chi_{\mathrm{ns}}^2N_aN_{\mathrm{ph}} \\[5pt]
\mathrm{Cov}(&\hat{J}_z(\tau),\hat{S}_b(\tau)) = \mathrm{Cov}(\hat{S}_b(\tau),\hat{J}_z(\tau)) \approx  \chi_{\mathrm{ns}}N_a\sqrt{N_{\mathrm{ph}}}
\end{align}
where here $N_{\mathrm{ph}} = \beta_0^2\tau$. Also, we have defined $\chi_{\mathrm{ns}} \equiv \frac{g^2}{c\Delta}$, where the subscript denotes no spontaneous emission. We finally find the quantum-enhancement parameter:

\begin{align}
\xi_{s_2}^{\mathrm{ns}}(\tau) \approx e^{\chi_{\mathrm{ns}}^2N_{\mathrm{ph}}}\left(1 - \frac{\chi_{\mathrm{ns}}^2N_{\mathrm{ph}}N_a}{\chi_{\mathrm{ns}}^2N_{\mathrm{ph}}N_a + 1/2}\right)^{1/2} \, .
\label{ksi_s_nse}
\end{align}

By inspection of Eq.~(\ref{ksi_s_nse}) we see that the parameters that affect the sensitivity of our signal are the total number of photons $N_{\mathrm{ph}}$, the quantization area of the light field $A$ (through $g$), the detuning $\Delta$, and the total number of atoms $N_a$. We also notice that we can always increase the sensitivity of our signal by just increasing $\chi_{\mathrm{ns}}N_{\mathrm{ph}}N_a$ up to a point that the increase of $e^{\chi_{\mathrm{ns}}^2N_{\mathrm{ph}}}$ becomes dominant. This is essentially the point that $\langle J_x \rangle$ (denominator of Eq.~(\ref{signal_2})) has decreased so much that the sensitivity starts decaying. Following that strategy we can always achieve better sensitivity than the standard quantum limit (SQL), as seen in Fig.[\ref{no_spont_emis}]. Here, we find the minimum of $\xi_{s_2}^{\mathrm{ns}}$ by taking the derivative with respect to the collective parameter $\nu = \chi_{\mathrm{ns}}^2N_{\mathrm{ph}}$:

\begin{align}
(\xi_{s_2}^{\mathrm{ns}})_\mathrm{min} = \sqrt{\frac{e}{N_a}} \, .
\end{align}
We see that the minimum depends on the inverse of the number of atoms, while the optimum number of photons for which we take that minimum is
\begin{align}
N_{\mathrm{ph}}^{\mathrm{opt}} = \frac{1}{2\chi_{\mathrm{ns}}^2} \, .
\end{align}

\begin{figure}[h]
	\begin{center}
		\includegraphics[width=1.1\columnwidth]{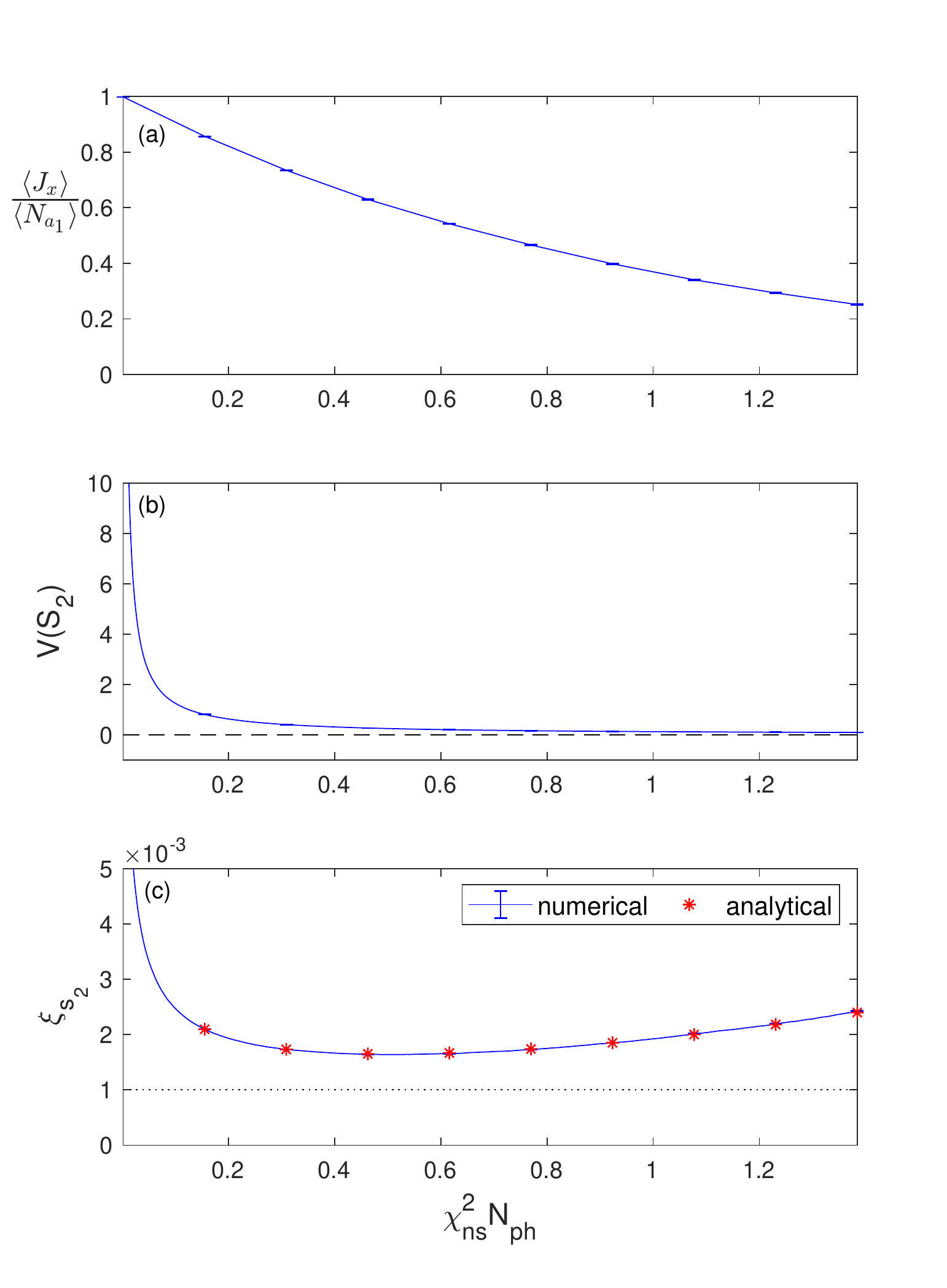}
		\caption{(a) $\langle J_x\rangle/\langle N_{a_1}\rangle$ (b) $\mathrm{Var}(S_2)$ and (c) $\xi_{s_2}$ with respect to the collective parameter $\nu = \chi_{\mathrm{ns}}^2N_{\mathrm{ph}}$. (a): The decay is due to over-squeezing the state, since we do not consider spontaneous emission here. This causes the squeezing parameter to reach a minimum value (c). The black dashed line in (b), points to zero, just to reassure the $\mathrm{Var}(S_2)$ is always positive. In (c) the black dotted line represents the Heisenberg limit. The parameter values are $A=10^{-10} \, \mathrm{m}^2$, $\Delta=10^{2} \, \mathrm{GHz}$, $N_{a}=10^6$. The error bars are barely distinguishable from all lines.}
		\label{no_spont_emis}
	\end{center}
\end{figure}

\subsection{Spontaneous emission}\label{Anal_se}
With the inclusion of spontaneous emission ($\gamma_3 > 0$), the calculation of the atomic expectation values is much more complicated. We begin by ignoring the effect that quantum fluctuation in the optical field has on the spontaneous emission. That is
\begin{equation}
e^{-g^2 \Gamma \int_0^t \bhatd(z,t^\prime)\bhat(z,t^\prime) dt^\prime} \approx e^{-g^2 \Gamma \beta_0^2 t}
\end{equation}
such that
\begin{align}
\langle\hat{N}_{a_1}(t)\rangle \approx \frac{N_a}{2}\epsilon(t)
\end{align}
where $\epsilon(t) \equiv e^{-2g^2\Gamma \beta_0^2 t}$ indicates how fast we lose atoms from our system. Following the same strategy as before we find
\begin{align}
&\mathrm{Var}(\hat{S}_b(\tau)) \approx 2 + 4\chi_1^2N_{\mathrm{ph}}N_a\overline{\epsilon(\tau)}  \, ,
\end{align}
and
\begin{align}
\mathrm{Cov}(\hat{J}_z(\tau),&\hat{S}_b(\tau)) = \mathrm{Cov}(\hat{S}_b(\tau),\hat{J}_z(\tau)) \approx \chi_1\sqrt{N_{\mathrm{ph}}}N_a\epsilon(\tau)\, ,
\end{align}
where we have defined $\chi_1\equiv \frac{g^2\Omega}{c}$ and $\overline{\epsilon(\tau)} = \frac{1}{\tau}\int_{0}^{\tau}\epsilon(t)dt$ which is the time average of the decay. Note that $\chi_1 = \chi_{\mathrm{ns}}$ in the no spontaneous emission case ($\gamma_3=0$).  The spin-squeezing parameter is therefore
\begin{align}
\xi_{s_2} \approx e^{(\chi_1^2 + \chi_2)N_{\mathrm{ph}}}\left(1 - \frac{\chi_1^2N_{\mathrm{ph}}N_a\epsilon(\tau)}{\chi_1^2N_{\mathrm{ph}}N_a\overline{\epsilon(\tau)} + 1/2}\right)^{1/2} \, ,
\label{ksi_se_paper}
\end{align} 
where we have defined $\chi_2 \equiv \frac{g^2\Gamma}{c}$ and now the decay factor can be expressed as $\epsilon(\tau) = e^{-2\chi_2N_{\mathrm{ph}}}$. We also find for the time average of the decay factor that $\overline{\epsilon(\tau)} = \frac{1 - \epsilon(\tau)}{2\chi_2 N_{\mathrm{ph}}}$. 

 By inspecting Eq.~(\ref{ksi_se_paper}) it is clear that the case with spontaneous emission is more complicated. We notice again that we can increase the sensitivity by increasing the term $\chi_{1}^2N_{\mathrm{ph}}N_a\propto \frac{N_{\mathrm{ph}}N_a}{A^2\Delta^2}$ (for $\Delta>>\gamma_3$), but now we are restricted by the atomic loss rate $\epsilon = \exp{(-2\chi_2N_{\mathrm{ph}})}\propto \exp{\left(\frac{N_{\mathrm{ph}}}{A\Delta^2}\right)}$ (for $\Delta>>\gamma_3$). Hence, we have to find the appropriate parameter regime that balances between spin squeezing and atomic loss.
 
 \begin{figure}[h]
 	\begin{center}
 		\includegraphics[width=1.1\columnwidth]{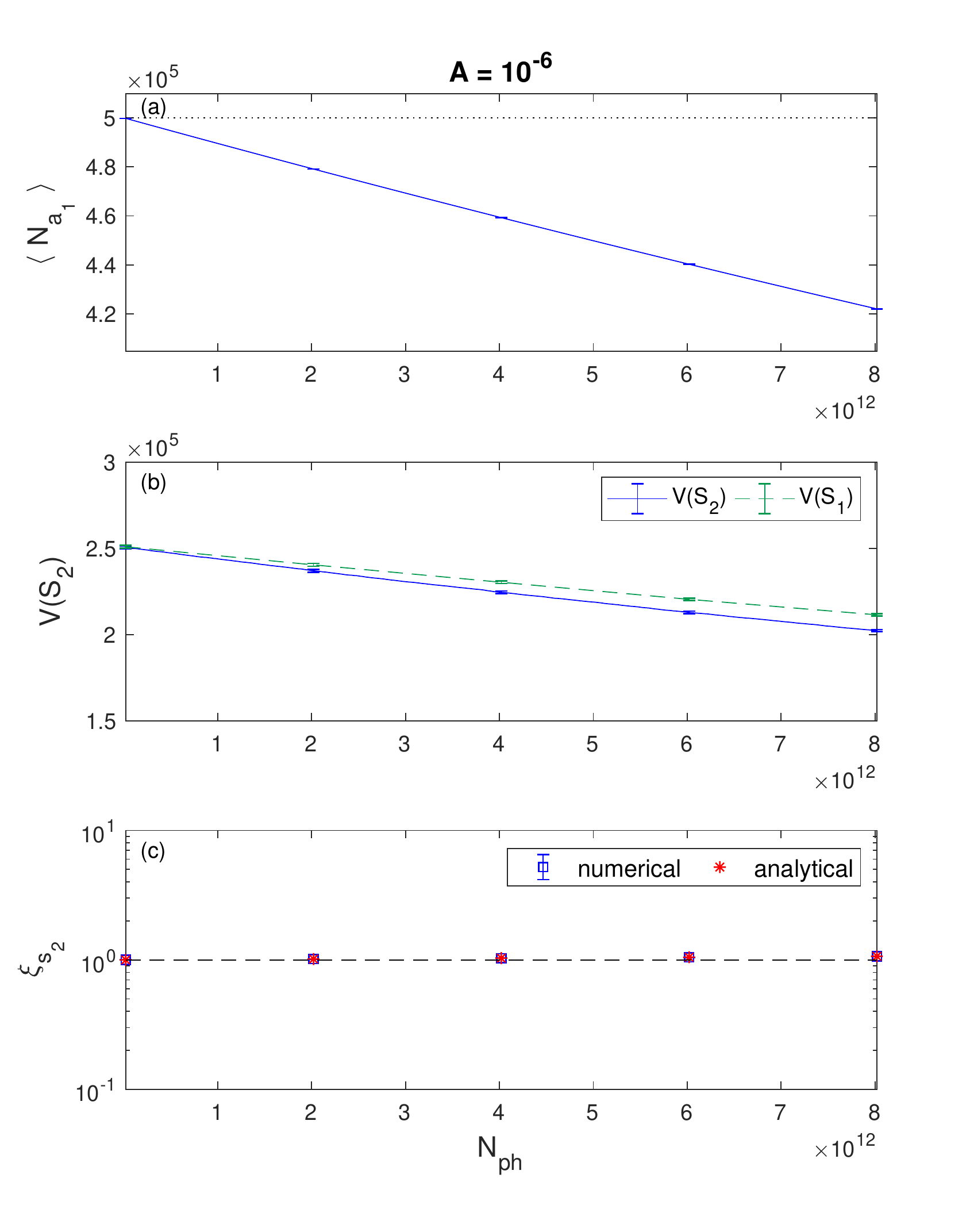}
 		\caption{(a) $\langle N_{a_1}\rangle$, (b) $\mathrm{Var}(S_1)$ (green dashed line) and $\mathrm{Var}(S_2)$ (blue solid line) (c) $\xi_{s_2}$ numerical (blue squares) and analytical (red asterisks) with respect to number of photons. In (a) the black dotted line shows the initial atomic population, while the black dashed line in (c) represents the SNL. The parameter values are $A=10^{-6} \, \mathrm{m}^2$, $\Delta=10^{2} \, \mathrm{GHz}$, $N_{a}=10^6$.}
 		\label{spont_emis_A=10^{-6}}
 	\end{center}
 \end{figure}

 We present simulations of our analytical results for $\xi_{s_2}$, Fig.~[\ref{spont_emis_A=10^{-6}}(c)]-[\ref{spont_emis_A=10^{-10}}(c)], for three different quantization area values, $A = \left(10^{-3} \,\mathrm{m}\right)^2$, $A =\left(10^{-4} \,\mathrm{m}\right)^2$ and $A = \left(10^{-5} \,\mathrm{m}\right)^2$. For each different area value we essentially change the number of photons and detuning appropriately in order to obtain best sensitivities . For $A=\left(10^{-3} \,\mathrm{m}\right)^2$ we notice that we never obtain enhanced sensitivity (compared to SQL) since the loss of atoms exceeds the resulting squeezing, Fig.\ref{spont_emis_A=10^{-6}} (c). As we decrease $A$ the atom-light interaction strengthens, increasing the sensitivity of our signal Fig.[\ref{spont_emis_A=10^{-8}},\ref{spont_emis_A=10^{-10}}]. 
 
 \begin{figure}[h]
 	\begin{center}
 		\includegraphics[width=1.1\columnwidth]{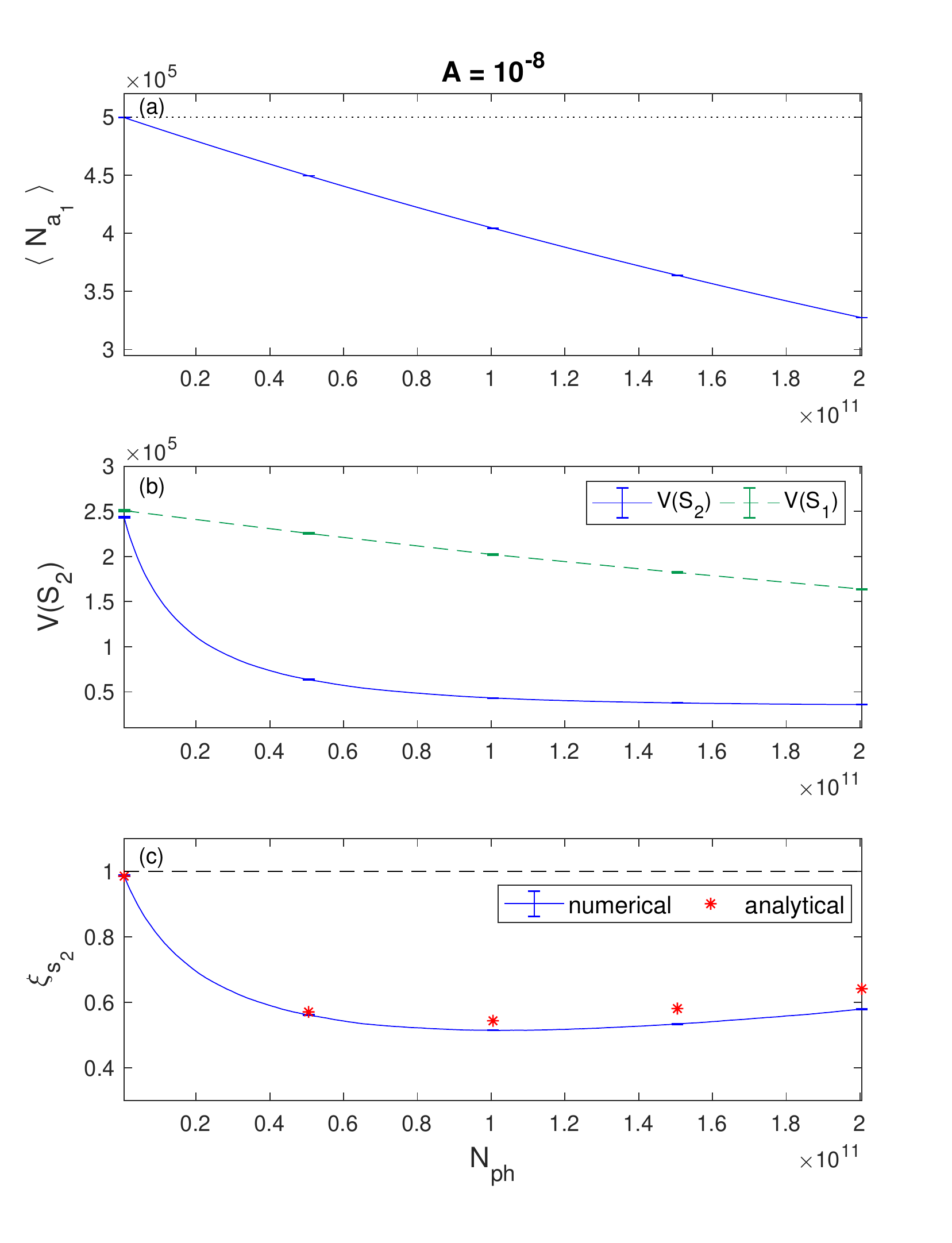}
 		\caption{(a) $\langle N_{a_1}\rangle$, (b) $\mathrm{Var}(S_1)$ (green dashed line) and $\mathrm{Var}(S_2)$ (blue solid line) (c) $\xi_{s_2}$ numerical (blue solid line) and analytical (red asterisks) with respect to number of photons. In (a) the black dotted line shows the initial atomic population, while the black dashed line in (c) represents the SNL. The parameter values are $A=10^{-8} \, \mathrm{m}^2$, $\Delta=10^{2} \mathrm{GHz}$, $N_{a}=10^6$.}
 		\label{spont_emis_A=10^{-8}}
 	\end{center}
 \end{figure}

\begin{figure}[h]
	\begin{center}
		\includegraphics[width=1.1\columnwidth]{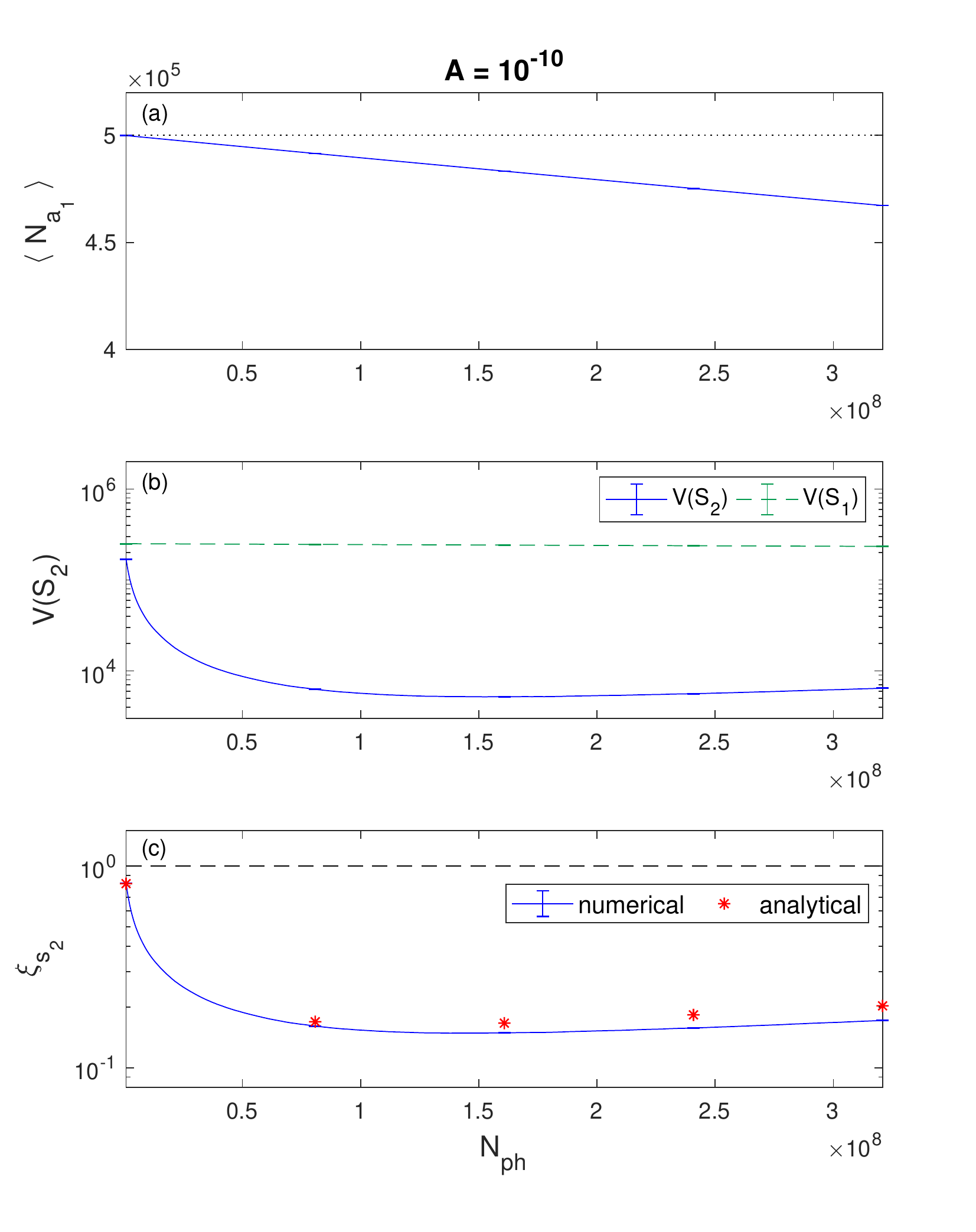}
		\caption{(a) $\langle N_{a_1}\rangle$, (b) $\mathrm{Var}(S_1)$ (green dashed line) and $\mathrm{Var}(S_2)$ (blue solid line) (c) $\xi_{s_2}$ numerical (blue solid line) and analytical (red asterisks) with respect to number of photons. In (a) the black dotted line shows the initial atomic population, while the black dashed line in (c) represents the SNL. The parameter values are $A=10^{-10} \, \mathrm{m}^2$, $\Delta=10^{2} \, \mathrm{GHz}$, $N_{a}=10^6$.}
		\label{spont_emis_A=10^{-10}}
	\end{center}
\end{figure}

In order to find the minimum of $\xi_{s_2}$, we express Eq.~(\ref{ksi_se_paper}) in terms of the dimensionless parameters $\mu \equiv \frac{\chi_1^2}{\chi_2} = \frac{g^2\Omega^2}{c\Gamma}$, $\lambda \equiv \chi_2N_{\mathrm{ph}}$ and $\zeta\equiv N_a \mu$. Hence, we can now write $\xi_{s_2}$ as
\begin{align}
\xi_{s_2} = e^{\lambda(1+\mu)}\left(1 - \frac{\zeta\epsilon(\tau)}{\zeta - \zeta\epsilon(\tau) + 1}\right)^{1/2} , 
\end{align}
where the decay can now be expressed as $\epsilon(\tau) = e^{-2\lambda}$. We work in a parameter regime where $\mu<<1$, such that
\begin{align}
\xi_{s_2} \approx e^{\lambda}\left(1 - \frac{2\zeta\lambda e^{-2\lambda}}{1 + \zeta - \zeta e^{-2\lambda}}\right)^{1/2} \, .
\label{ksi_opt}
\end{align}

In order to simplify things further, we consider the case where $\Delta>>\gamma_e$. In that case $\Omega\rightarrow \frac{1}{\Delta}$ and $\Gamma\rightarrow\frac{\gamma_3}{2\Delta^2}$, thus $\mu\rightarrow\frac{2g^2}{c\gamma_3}$. That means that $\mu$ only depends on the atomic properties and the quantization area of the light $A$ (through $g$) and consequently $\zeta \rightarrow\frac{2g^2}{c\gamma_3}N_a$. On the other hand $\lambda \rightarrow\frac{g^2\gamma_3}{2c}\frac{N_{\mathrm{ph}}}{\Delta^2}$ for $\Delta>>\gamma_3$. Hence, if we fix the value of $\zeta$, by choosing a specific value for the number of atoms $N_a$ and the area $A$, we only need to optimize $\xi_{s_2}$ with respect to $\lambda$ which is proportional to $N_{\mathrm{ph}}/\Delta^2$ in the regime $\Delta>>\gamma_3$. In Fig.~[\ref{ksi_s_optimised}] we followed that procedure for several different values of $\zeta$ and found the minimum of $\xi_{s_2}$ with respect to $\lambda$ using Eq.~(\ref{ksi_opt}). We notice that the sensitivity increases as we increase $\zeta$, which means either increasing $N_a$ or decreasing the area. Just to clarify here that by decreasing the area we also increase the atomic loss rate, which leads to loss of sensitivity. In that case we should also change the other parameters ($N_{\mathrm{ph}}/\Delta^2$) in order to counteract that effect, resulting at the end in better sensitivities. On the other hand, the increase of $N_a$ does not affect the loss rate of atoms and it solely improves the sensitivity. 

\begin{figure}[h]
	\begin{center}
		\includegraphics[width=1.1\columnwidth]{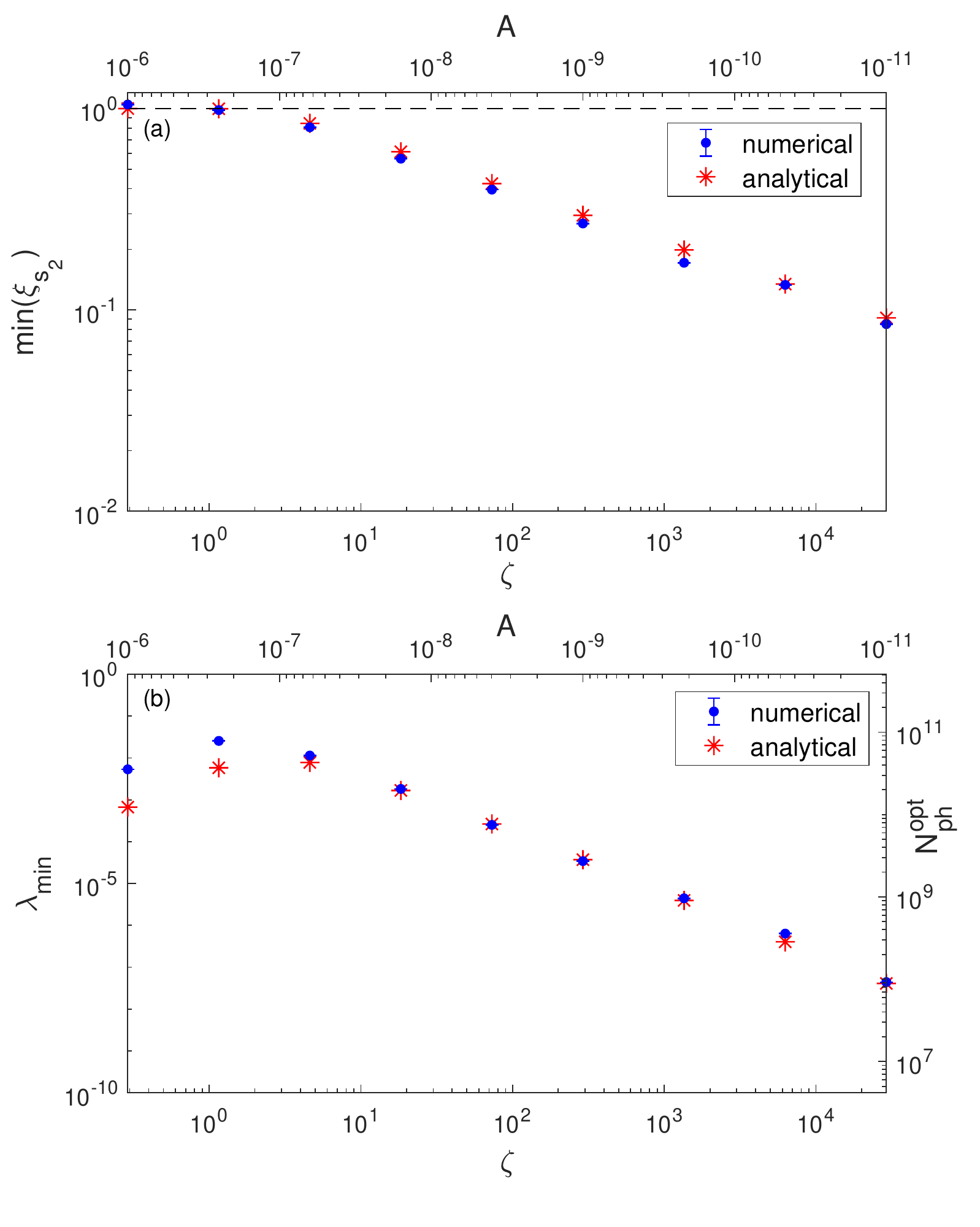}
		\caption{ (a) Minimum value of $\xi_{s_2}$ with respect to $\zeta$ (bottom x-axis) and $A$ (top horizontal-axis), (b) optimum $\lambda$ (left vertical-axis) and optimum number of photons $N_{\mathrm{ph}}^{\mathrm{opt}}$ (right y-axis) with respect to $\zeta$. In (a) the black dashed line represents the SNL.}
		\label{ksi_s_optimised}
	\end{center}
\end{figure}

\section{Numerical Solutions}\label{numerical}
We can solve for the dynamics of the system numerically by using the Truncated Wigner (TW) method \cite{quantum_noise_book_Gardiner}. From the Heisenberg equations of motion we can move to Fokker-Plank equations (FPEs) by using correspondences between quantum operators and Wigner variables. After truncating third and higher order terms we can map the FPEs into stochastic differential equations (SDE) which can be solved numerically with respect to the Wigner variables. We make the following correspondences $\hat{a}_1(t) \rightarrow \alpha_1(t)$, $\hat{b}_1(z,t)\rightarrow\beta_1(z,t)$ and $\hat{q}_{1_{i\mathrm{n}}}(t)\rightarrow\mathfrak{q}_{in}(t)$. We also consider the initial conditions $\alpha_1(0)=\alpha_{10} + \eta_1$,  $\beta_{01}(t) =\beta_0 + w_{b_1}(t)$ and $\mathfrak{q}_{in}(t) = w_{q_1}(t)$. $\eta_1$ is complex Gaussian noise satisfying $\overline{\eta_1}=0$ and $\overline{\eta_1^*\eta_1}=\frac{1}{2}$,  $w_{x}(t)$ is a complex Wiener noise satisfying $\overline{w_{x}(t)}=0$ where $x = b_1, \, q_1$. Also, $\overline{w_{b_1}(t)w_{b_1}(t')}=\frac{1}{2c}\delta(t-t')$ and $\overline{w_{q_1}(t)w_{q_1}(t')}=\frac{1}{2}\delta(t-t')$, where the bar represents averaging with respect to a large number of stochastic trajectories.

We consider the D2 transition line of $^{87}\mathrm{Rb}$ $(5^2S_{1/2}\rightarrow 5^2P_{3/2})$ for both atomic transitions, where the transition frequency is $\omega_{13}=\omega_{24}=\omega_a=2\pi c/\lambda$ and $\lambda = 780 \,\mathrm{nm}$. The spontaneous emission rate of the exited state is $\gamma_3 = \gamma_4 = 38.11 \, \mathrm{MHz}$ \cite{Steck:2015}.


In Fig.~[\ref{spont_emis_A=10^{-6}}]-[\ref{ksi_s_optimised}], we present the numerical simulations corresponding to the analytical results analysed in the previous section. We notice that our analytical and numerical results have almost perfect agreement, indicating that the approximations we made through the derivations do not have any significant effect in the final results.

\section{Squeezed Light}\label{sec7}
Up to this point we have only considered classical light sources. That is, we have assumed that the incoming light is a coherent state, with $\mathrm{Var}(\hat{Y}_{1_{\mathrm{in}}}) = 1$.  It is possible to increase the sensitivity of our final signal by considering a squeezed incoming light, where $\mathrm{Var}(\hat{Y}_{1_{\mathrm{in}}})_{\mathrm{sq}} = e^{-2r}$ and $r$ is the squeezing factor \cite{Bachor:2004}. In that case our analytical calculation for the spontaneous emission case results in
\begin{align}
\mathrm{Var}(\hat{S}_b)_{\mathrm{sq}} \approx 2\mathrm{Var}(\hat{Y}_1(\tau))_{\mathrm{sq}} \approx 2e^{-2r} + 4\chi_{\mathrm{ns}}^2N_aN_{\mathrm{ph}}
\end{align}
while the covariances remain the same. Hence, the quantum enhancement parameter become
\begin{align}
\xi_{s_2} \approx e^{(\chi_1^2 + \chi_2)N_{\mathrm{ph}}}\left(1 - \frac{\chi_1^2N_{\mathrm{ph}}N_a\epsilon(\tau)}{\chi_1^2N_{\mathrm{ph}}N_a\overline{\epsilon(\tau)} + e^{-2r}/2}\right)^{1/2} \, .
\label{ksi_squeezed_light_analytical}
\end{align} 
In Fig.~[\ref{squeezed_light_all_areas}] we notice that we obtain better sensitivity for all three area values compared to the coherent incident light (Fig.~[\ref{spont_emis_A=10^{-6}} - \ref{spont_emis_A=10^{-10}}]). In Fig.~[\ref{min_ksi_improvement_dB}] we show the numerical and analytical $\min(\xi_{s_2})$ for the three different area values, with respect to the degree of optical squeezing in the incoming light, $\mathcal{S}$, defined by
\begin{equation}
\mathcal{S} = 10\log\left( \frac{\sqrt{\mathrm{Var}(\tilde{Y}_{b_1})}}{\sqrt{\mathrm{Var}(Y_{b_1})}}\right) \, \mathrm{dB} \, ,
\end{equation}
where $\mathrm{Var}(Y_{b_1}) =1$ is the variance for a coherent state, and $\mathrm{Var}(\tilde{Y}_{b_1}) = e^{-2r}$, where $r$ is the squeezing factor.
Using squeezed incoming light gives an exponential rate of decrease for $\xi_s$ for all cases (for $A = 10^{-6}$ that holds for  $\gtrapprox 5 \mathrm{dB}$). In addition, for a light field with improvement $\gtrapprox 5 \mathrm{dB}$ we see that we can surpass the SNL even for the $A = 10^{-6}\mathrm{m^2}$ case, while that was impossible when we used a coherent initial state for the light field, Fig.~[\ref{spont_emis_A=10^{-6}}] (c). Finally, we notice in Fig~[\ref{min_ksi_improvement_dB}] that our analytical approximative model (red stars) given by Eq.~(\ref{ksi_squeezed_light_analytical}) agrees well with our numerical results (blue circles).




\begin{figure}[h]
	\begin{center}
		\includegraphics[width=1.1\columnwidth]{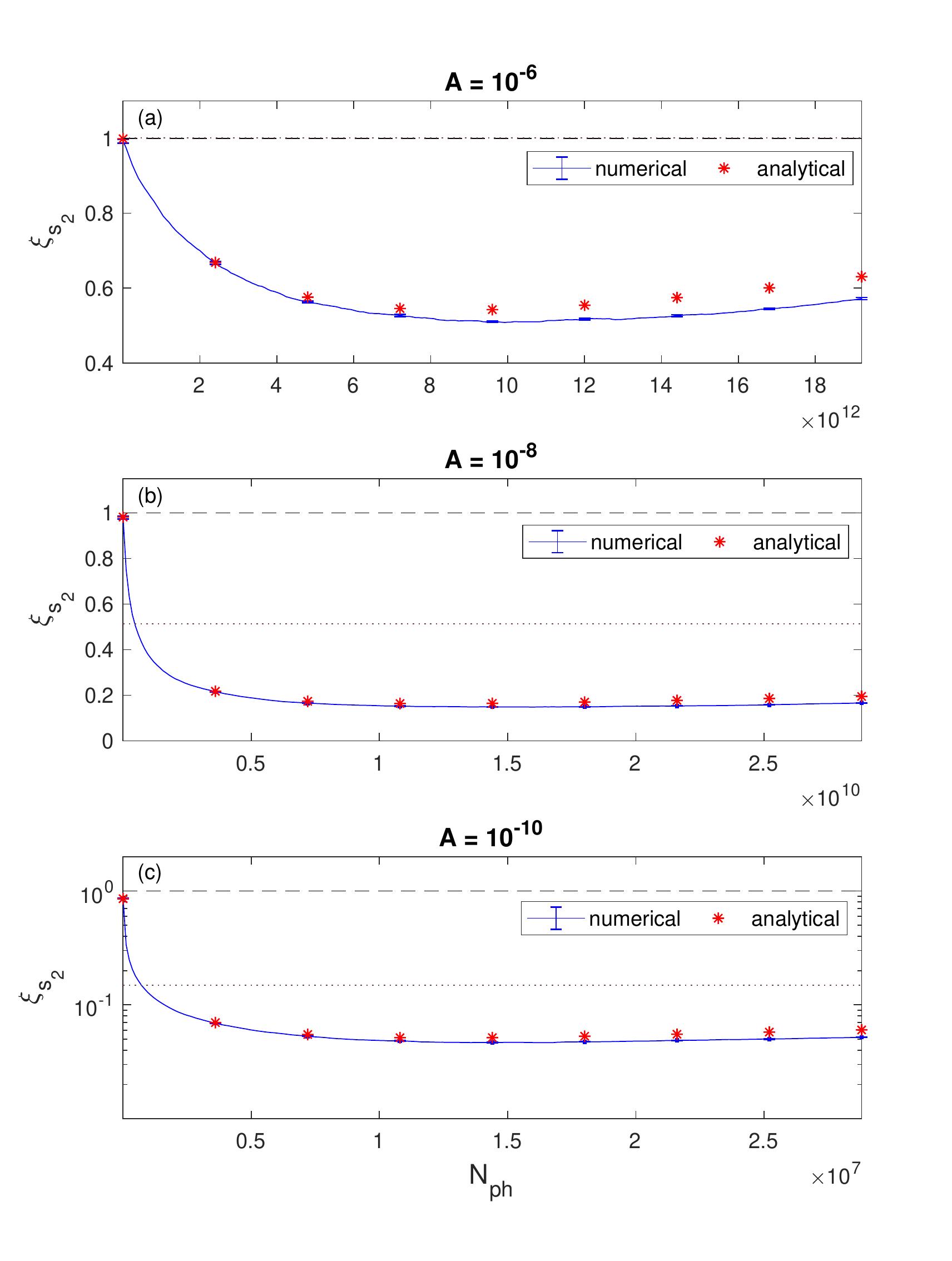}
		\caption{ We consider squeezed incoming light and we examine the numerical (blue solid line) and analytical (red asterisks) evolution of $\xi_{s_2}$  with respect to the number of photons for all three area values. The brown dotted lines show the $\mathrm{min}(\xi_{s_2})$ of the corresponding cases in Fig.~[\ref{spont_emis_A=10^{-6}}] - [\ref{spont_emis_A=10^{-10}}]. The other parameter values are $r = \ln 10$, $\Delta=10^{2} \, \mathrm{GHz}$, $N_{a}=10^6$.}
		\label{squeezed_light_all_areas}
	\end{center}
\end{figure}

\begin{figure}[h]
	\begin{center}
		\includegraphics[width=1.0\columnwidth]{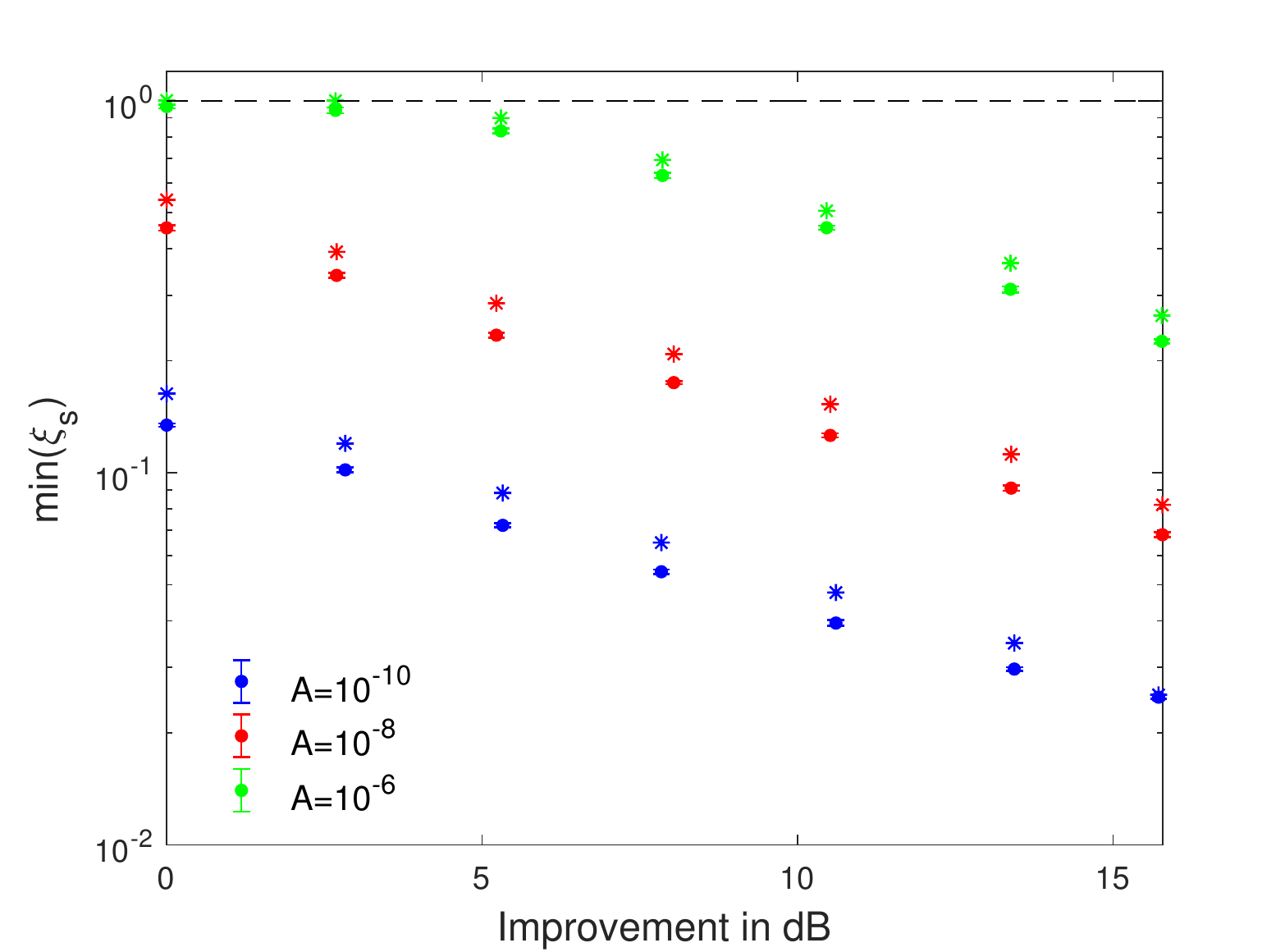}
		\caption{Analytical (stars) and numerical (circles) calculation of the minimum value of $\xi_{s_2}$ with respect to the improvement in dB of the incoming light field, for the three different area values. The black dashed line represents the SNL. The other parameter values are $\Delta=10^{2}$ GHz, $N_{a}=10^6$. }
		\label{min_ksi_improvement_dB}
	\end{center}
\end{figure}

\section{Cavity Dynamics}\label{sec8}
\begin{figure}[h]
	\begin{center}
		\includegraphics[width=0.9\columnwidth]{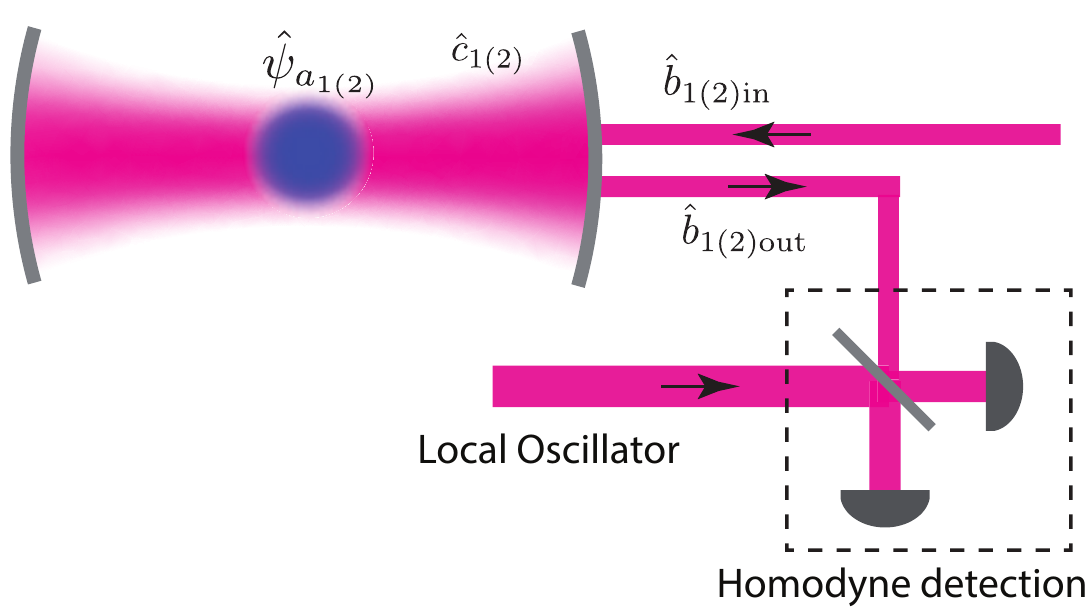}
		\caption{QND interaction boosted by an optical cavity. After interacting with the atomic ensemble, the light exiting the cavity $\bhat_{1(2) \mathrm{out}}$ is measured via homodyne detection.}
		\label{cavity_scheme}
	\end{center}
\end{figure}
We can further boost the sensitivity of our signal with the addition of an optical cavity, as it essentially increases the atom-light coupling Fig.~[\ref{cavity_scheme}]. We consider a dual-frequency cavity with resonant frequencies $\omega_{c_1}$ and $\omega_{c_2}$ detuned from the two atomic transitions $|1\rangle\rightarrow |3\rangle$ and $|2\rangle\rightarrow |4\rangle$ by detunings $\Delta_1$ and $\Delta_2$ respectively. In the Hamiltonian of our system, Eq.~(\ref{Hamilt}) we interchange the continuous light field annihilation operators $\hat{b}_1(z,t)$ and $\hat{b}_2(z,t)$ with the cavity mode annihilation operator $\hat{c}_1$ and $\hat{c}_2$, giving
\begin{align}
\hat{H}_{\mathrm{tot}}^c &= \hbar\omega_{c_1}\hat{c}_1^\dagger\hat{c}_1 + \hbar\omega_{c_2}\hat{c}_2^\dagger\hat{c}_2 \nonumber\\  & +\hbar\int_{-\infty}^{\infty}dz\left(\omega_{13}\hat{\psi}_3^{\dagger}(z,t)\hat{\psi}_3(z,t) + \omega_{24}\hat{\psi}_4^{\dagger}(z,t)\hat{\psi}_4(z,t)\right) \nonumber\\
& + \; \hbar g_{c_1}\int_{-\infty}^{\infty}\left(\hat{\psi}_1^{\dagger}(z,t)\hat{\psi}_3(z,t)\hat{c}_1^\dagger(t) + \mathrm{h.c}\right)dz \nonumber\\ 
& + \; \hbar g_{c_2}\int_{-\infty}^{\infty}\left(\hat{\psi}_2^{\dagger}(z,t)\hat{\psi}_4(z,t)\hat{c}_2^\dagger(t) + \mathrm{h.c}\right)dz 
\label{Hamilt_cac}
\end{align}
The coupling strength constants are defined as $g_{c_1} =\frac{d_{13}}{\hbar}\left(\frac{\hbar\omega_{c_1}}{2\epsilon_0 V}\right)^{1/2}$ and $g_{c_2} =\frac{d_{24}}{\hbar}\left(\frac{\hbar\omega_{c_2}}{2\epsilon_0 V}\right)^{1/2}$ where $V = AL$ is the volume of the cavity, $A$ is the light quantization transverse area and $L$ is the cavity length. Using the standard input output formalism \cite{Walls:2008} we obtain the equation of motion for $\hat{c}_1$
\begin{align}
\partial_t\hat{c}_1 = -\frac{i}{\hbar}\left[\hat{c}_1,\hat{H}_{\mathrm{tot}}^c\right] - \frac{\kappa}{2}\hat{c}_1 + \sqrt{\kappa}\hat{b}_{1_{\mathrm{in}}}(t) \, ,
\end{align}
where $\kappa$ is the cavity photon decay rate, and $\hat{b}_{1_{\mathrm{in}}}(t) = \sqrt{c}\hat{b}_1(z_L,t)$ where $c$ is the speed of light and $\hat{b}_1(z_L,t)$ is the continuous in space annihilation operator of the incoming light field used in the previous sections, satisfying $[\hat{b}_{1_{\mathrm{in}}}(t),\hat{b}_{1_{\mathrm{in}}}(t')] =\delta(t-t')$. Another important quantity is the light field leaking out of the cavity
\begin{align}
\hat{b}_{1_{\mathrm{out}}}(t) = \sqrt{\kappa}\hat{c}_1(t) - \hat{b}_{1_{\mathrm{in}}}(t) \, .
\label{b1_out}
\end{align}
In this case, $\hat{b}_{1_{\mathrm{in}}}(t)$ is an input light field that coherently drives the dynamics of the cavity, but now the mode of the cavity, $\hat{c}_1$, is the one that interacts with the atomic ensemble and is entangled with the atomic ground-state number operator. Again, we incorporate spontaneous emission following the same method as in Sec.~(\ref{scheme_sec}), i.e we use Eq.~(\ref{Lang_eq_psi3}) in order to eliminate $\hat{\psi}_3(z,t)$ from the equations of motion for $\hat{\psi}_1(z,t)$ and $\hat{c}_1$. After making the single mode approximation for $\hat{\psi}_1(z,t)$ and $\hat{d}_{1_\mathrm{in}}(z,t)$ using again the same mode functions for both of them, and moving to a rotating frame with respect to the cavity resonance frequency we obtain
\begin{subequations}
\begin{align}
	&\begin{aligned}
		\partial_t\hat{a}_1(t) = \; ig^2(\Omega + &i\Gamma)\tilde{c}_1^\dagger(t)\tilde{c}_1(t)\hat{a}_1(t) \, + \\ 
		&+ \, g_c\frac{\sqrt{\gamma_3}}{\Delta_1 - i\gamma_3/2}\tilde{c}_1^\dagger(t)\tilde{q}_{1_\mathrm{in}}(t)
	\end{aligned} \\[5pt]
	&\begin{aligned}
		\partial_t\tilde{c}_1(t) =& \left[ig^2(\Omega + i\Gamma)\hat{a}_1^\dagger\hat{a}_1 - \frac{\kappa}{2}\right]\tilde{c}_1(t) \, +  \\
		 &+ \, g_c\frac{\sqrt{\gamma_3}}{\Delta_1 - i\gamma_3/2}\hat{a}_1^\dagger(t)\tilde{q}_{1_\mathrm{in}}(t) + \sqrt{\kappa} \, \tilde{b}_{1_\mathrm{in}}(t)
	\end{aligned} 
\end{align}
\end{subequations}
where $\tilde{c}_1(t) = \hat{c}(t)e^{i\omega_{c_1}t}$, $\tilde{b}_{1_\mathrm{in}}(t) = \hat{b}_{1_\mathrm{in}}(t)e^{i\omega_{c_1}t}$, $\tilde{q}_{1_\mathrm{in}}(t) = \hat{q}_{1_\mathrm{in}}(t)e^{i\omega_{c_1}t}$.

%
%
%
%
%
%
%
To investigate the dynamics, we use the TW method, again making the appropriate correspondences, in order to numerically examine the dynamics of our system. In Fig.~[\ref{cavity_dynamics}] we plot the time evolution of the number of atoms and the number of cavity photons as well as the intensity of the input and output fields. We see that the cavity comes into its steady state after time $t \gg 1/\kappa$. As such, the rate of incoming photons should be larger than the rate of loss, ie $\langle  \hat{b}^\dagger_{1_{\mathrm{in}}} \hat{b}_{1_{\mathrm{in}}} \rangle \gg \kappa$, to ensure $\langle \hat{N}_{c_1}\rangle = \langle\hat{c}_1^\dagger\hat{c}_1\rangle \gg 1$. In our numerical simulations we have fixed the total interaction time $\tau = 10^{-4} >> 1/\kappa = 10^{-6}$ and we change the number of cavity photons, which is the parameter affecting the dynamics of our system, by just changing the intensity of the incoming light field $\langle \hat{b}_{1_{\mathrm{in}}}^\dag \hat{b}_{1_{\mathrm{in}}}\rangle$.

\begin{figure}[h]
	\begin{center}
		\includegraphics[width=1.1\columnwidth]{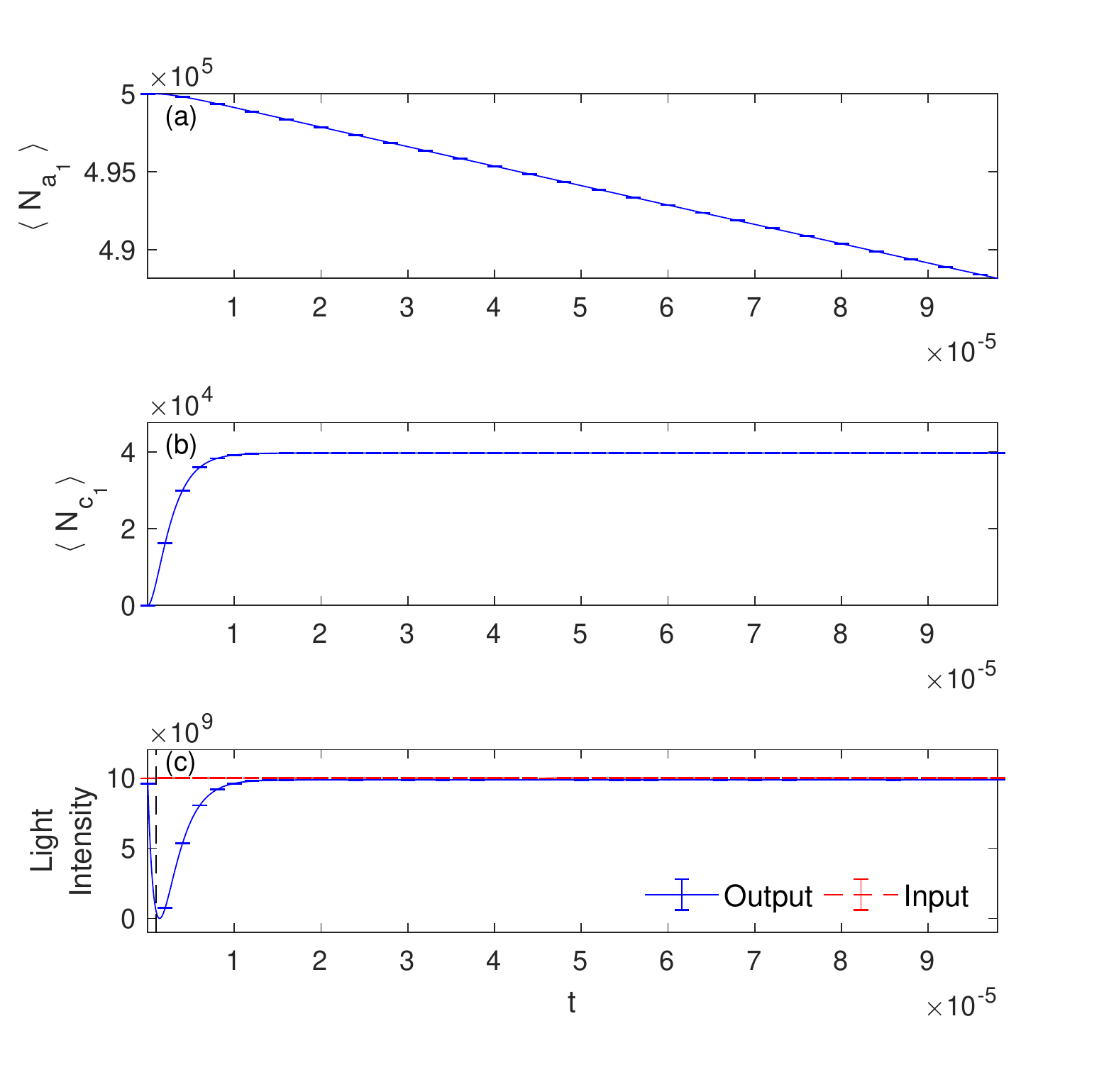}
		\caption{ Cavity dynamics: time evolution of (a) mean number of atoms in state $|1\rangle$, $\langle N_{a_1}\rangle$, (b) mean number of cavity photons $\langle N_{c_1} \rangle$, (c) Intensity of input light (red dashed line) and intensity of the leaking output light field from the cavity (blue line). The vertical black dashed line is drawn at the time point $1/\kappa$. We notice that we need $\tau \gg 1/\kappa$, in order to reach the cavity steady state. Other parameter values: $A = 10^{-8} \mathrm{m^2}$, $\Delta=10^{2} \, \mathrm{GHz}$, $N_{a}=10^6$ and $\kappa = 1 \, \mathrm{MHz}$.}
		\label{cavity_dynamics}
	\end{center}
\end{figure}

We measure a combined signal of the same form as in the free space case, but now we measure an observable of the output field, $\hat{b}_{1_{\mathrm{out}}}(t)$, since we do not have any direct access to the cavity mode. The output field contains information about atomic observables through Eq.~(\ref{b1_out}). Similarly with Sec.~(\ref{measurement_sec}) we use as our light observable the difference of the phase quadratures of a specific mode of the output fields.


We plot $\xi_{s_2}$ for the same area values as for Fig.~[\ref{spont_emis_A=10^{-6}}]-[\ref{spont_emis_A=10^{-10}}] with $\kappa = 10^6$ Hz. Here, we noticed that for $\Delta = 10^2\mathrm{GHz}$ and area values smaller than $A = 10^{-8} \mathrm{m^2}$ we have to decrease the incoming light intensity at a level that we tend to a regime where $\langle \hat{N}_{c_1} \rangle \rightarrow 1$. We can avoid that by just increasing appropriately the detuning $\Delta = 10^4\mathrm{GHz}$, in order to obtain the same interaction strength. Assuming a cavity of length $L = 10$ cm, this corresponds to a finesse of $\sim 10^4$. Our choice of cavity parameters is motivated by a cavity that could be added to an existing atom interferometry set-up, and can be installed outside the vacuum system. We use a range of different intensities for the incoming light field to determine the best sensitivity. Comparing Fig.~[\ref{spont_emis_A=10^{-6}}]-[\ref{spont_emis_A=10^{-10}}] with Fig.~[\ref{ksi_s_all_areas_cav}] it is apparent that we achieve better sensitivities by adding a cavity, than just using free space light fields. Although we don't have any analytical results for the case of the cavity, due to the complexity of that model, we examined numerically if the dynamics of  the system has the same behaviour as in the free space case. We concluded that we can find the optimum of the sensitivity using the same procedure as in Sec.~[\ref{numerical}]. Namely for a particular value of $A$ (or equivalently $V = AL$) and $N_a$ we can find the minimum of $\xi_{s_2}$ with respect to the remaining parameters $N_{c_1}/\Delta_1^2$. Here we have one parameter more, the photon decay rate from the cavity, $\kappa$. We notice that we have better sensitivities for smaller values of $\kappa$, thus for larger cavity quality factors (see Fig.~[\ref{cav_ksi_s_optimum}]). However, in the cavity case we are more constrained on the parameter values we could use, as they should satisfy $\langle \hat{b}^\dag_{1_{\mathrm{in}}} \hat{b}_{1_{\mathrm{in}}}\rangle > \kappa$ and $\tau > 1/\kappa$ as we discussed earlier. 

\begin{figure}[h]
	\begin{center}
		\includegraphics[width=1.1\columnwidth]{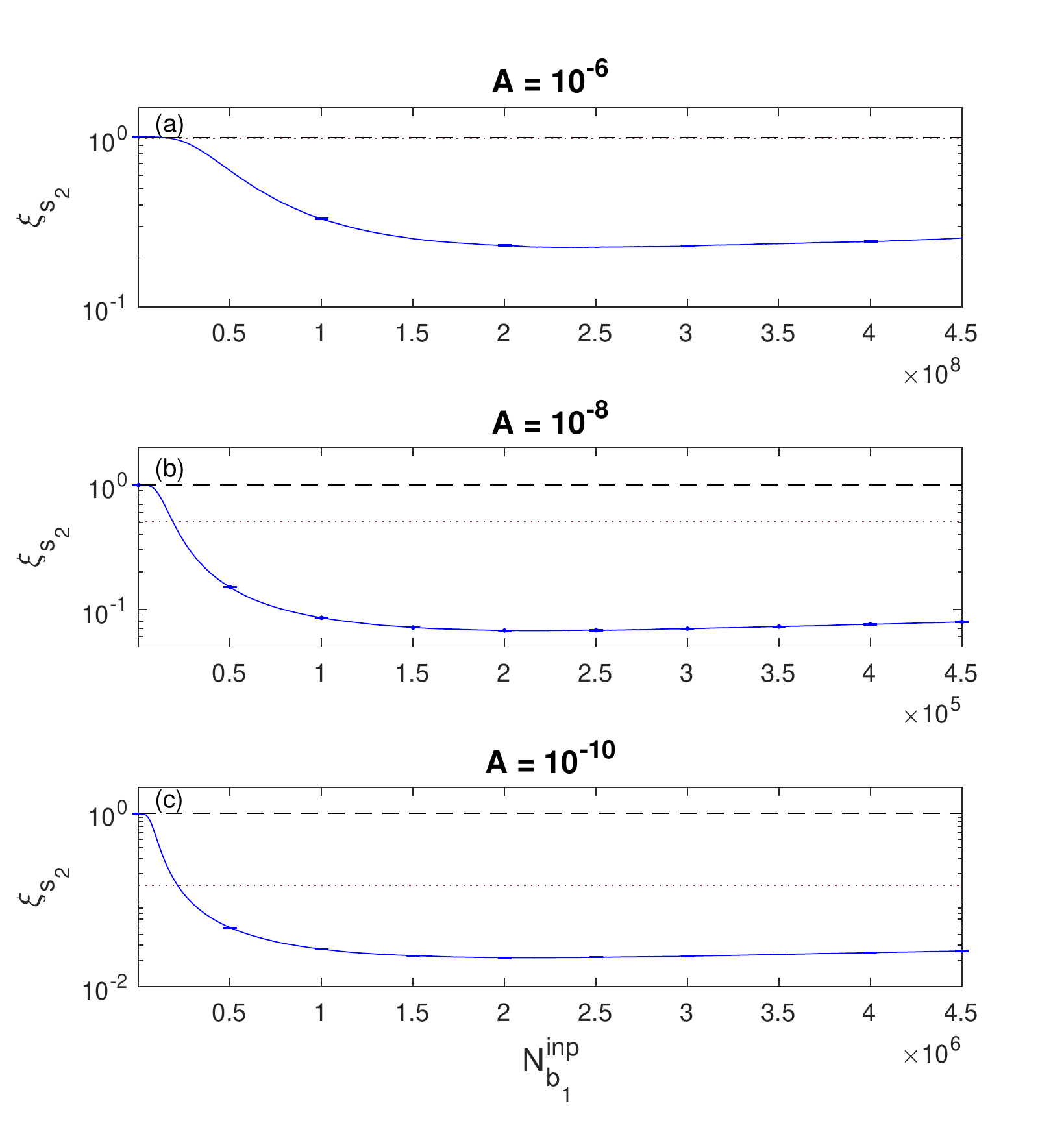}
		\caption{ Cavity: $\xi_s$ with respect to $N_{b_1}^{\mathrm{inp}}$ for different values of A.  The brown dotted lines show the $\mathrm{min}(\xi_{s_2})$ of the corresponding cases in Fig.~[\ref{spont_emis_A=10^{-6}}] - [\ref{spont_emis_A=10^{-10}}]. The parameter values are $N_{a}=10^6$ and $k=1 \, \mathrm{MHz}$ and $\Delta=10^{2} \, \mathrm{GHz}$, except (c) where we used $\Delta=10^{4} \, \mathrm{GHz}$, for the reasons discussed in the main text.}
		\label{ksi_s_all_areas_cav}
	\end{center}
\end{figure}

\begin{figure}[h]
	\begin{center}
		\includegraphics[width=1\columnwidth]{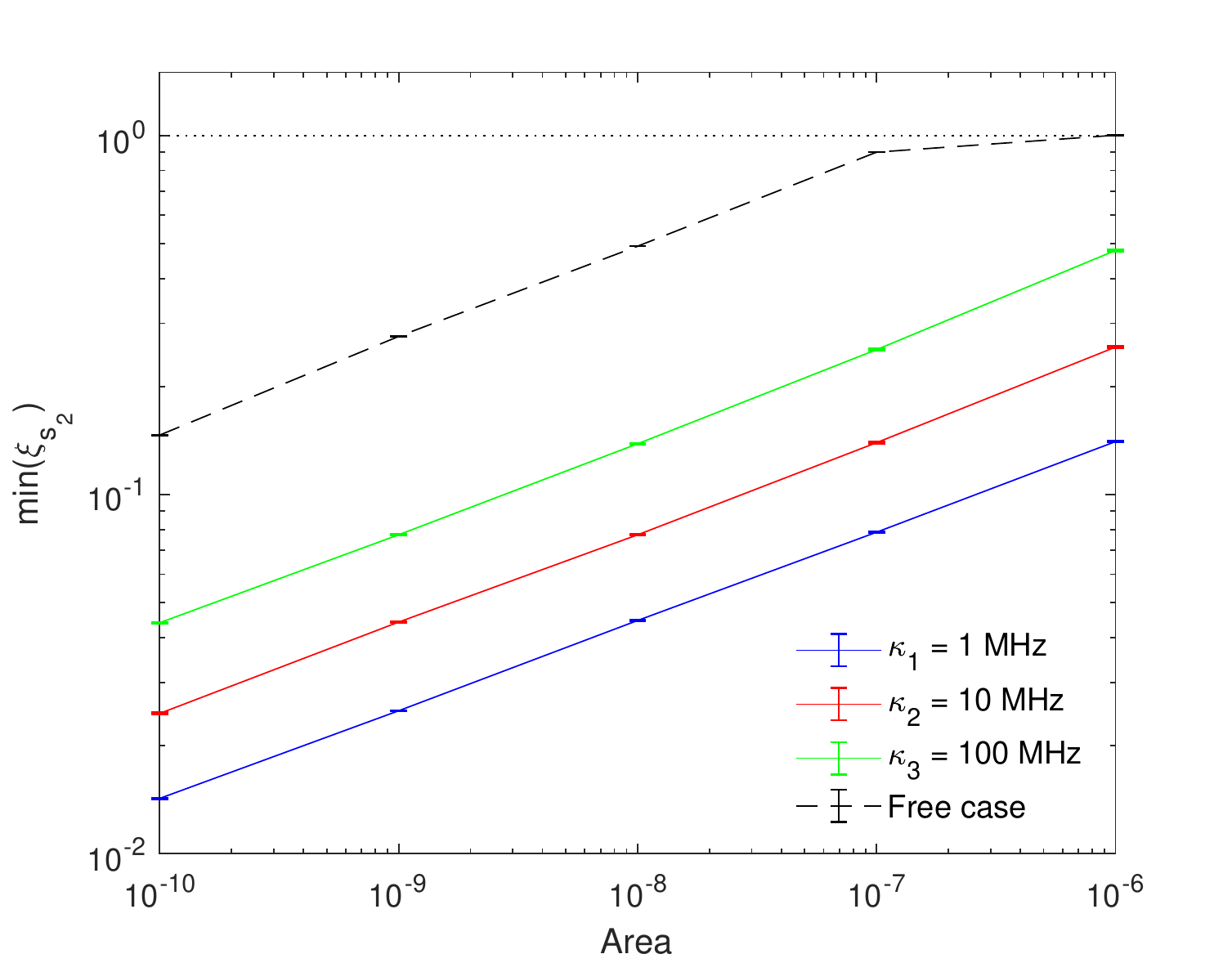}
		\caption{ Minimum value of $\xi_{s_2}$ with respect to the area, for three different values of $\kappa$ for the cavity case. We also plot the free space case (black dashed line). The black dotted line represents the SNL. The other parameter values are $N_{a}=10^6$ and $\Delta=10^{2} \, \mathrm{GHz}$, except for the area values $A = 10^{-9}\mathrm{m^2}$ and $A = 10^{-10} \mathrm{m^2}$ in all cavity lines where we used $\Delta = 10^4 \mathrm{GHz}$, for the reasons we mentioned in the main text.}
		\label{cav_ksi_s_optimum}
	\end{center}
\end{figure}

\section{Conclusion}
We have analysed the creation of spin-squeezing in an ensemble of Bose-condensed atoms via quantum non-demolition measurement, considering both freely propagating light, and optical cavities. We found that the determining factor in the quality of spin-squeezing produced was the cross-sectional area of the optical beam used to probe the spin of the atomic system, with small areas leading to higher atom-light coupling, and a larger phase shift on the light for a given level of spontaneous emission. Of course, varying the intensity, detuning, or duration of the incoming light also affects the level of spin squeezing. However, for a given area, fixing two of these parameters while adjusting the remaining one would always lead to the same optimum. For the D2 transition in $^{87}$Rb atoms, we found that for the case of freely propagating light, no squeezing was possible when the cross-sectional area of the atom-light interaction was larger than $\sim 10^{-6}$ m$^2$ due to loss of atoms due to spontaneous emission, regardless of the intensity or detuning of the incoming light. For areas less than this, we found significant spin-squeezing was possible, with an area of $10^{-11}$ m$^2$ leading to a spin squeezing value of $\sim 3\times 10^{-2}$, which corresponds to a potential improvement of atom interferometric sensitivity of $\sim 33$, which is equivalent to increasing the number of atoms by a factor of $1000$. The use of optical squeezing improved the level of quantum enhancement further, and relaxed the restrictions on the area of the light. Finally, we considered the use of an optical cavity. For reasonably achievable cavity parameters, we found approximately an order of magnitude increase over what was achievable in the free space case. 

\section{Acknowledgements}
The authors would like to acknowledge useful discussions with Barry Garraway, Stuart Szigetti, Joseph Hope, and John Close. M.K. and J.A.D. received funding from UK EPSRC through the Networked Quantum Information Technology (NQIT) Hub, Grant No. EP/M013243/1.

\bibliography{simon_QND,QND_bib}

\clearpage

\onecolumngrid
\appendix

\section{Introduction}\label{appendix_intro}

We consider the combined signal:

\begin{align}
\hat{S}_2(\tau) = \hat{J}_z(\tau) - \hat{J}_z^{\mathrm{inf}}(\tau)
\end{align} 

where 

\begin{align}
\hat{J}_z^{\mathrm{inf}}(\tau) = G\hat{S}_b(\tau),
\qquad
\hat{S}_b(\tau) = \hat{Y}_2(\tau) -\hat{Y}_1(\tau)
\end{align}

For simplicity in the following we will present the time dependence explicitly only in our final results or when it is considered necessary. The variance of $\hat{S}_2$ would be given by:

\begin{align}
\mathrm{Var}(\hat{S}_2) = \mathrm{Var}(\hat{J}_z) + G^2\mathrm{Var}(\hat{S}_b) - 2G\mathrm{Cov}(\hat{J_z},\hat{S}_b)
\label{Var_S2}
\end{align}

since $\mathrm{Cov}(\hat{J_z},\hat{S}_b) = \mathrm{Cov}(\hat{S}_b,\hat{J_z})$. We minimise $\mathrm{Var}(\hat{S}_2)$ with respect to $G$:

\begin{align}
G = \frac{\mathrm{Cov}(\hat{J_z},\hat{S}_b)}{\mathrm{Var}(\hat{S}_b)}
\label{gain}
\end{align}

Inserting that back in Eq.~(\ref{Var_S2}) we get:

\begin{align}
\mathrm{Var}(\hat{S}_2) =\mathrm{Var}(\hat{J}_z) - \frac{\mathrm{Cov}^2(\hat{J_z},\hat{S}_b)}{\mathrm{Var}(\hat{S}_b)}
\label{Var_S_2}
\end{align}

%
%
%

So, in order to calculate $\mathrm{Var}(\hat{S}_2)$ we need the covariance between $\hat{J}_z$ and $\hat{S}_b$, $\mathrm{Cov}(\hat{J_z},\hat{S}_b)$, and the variance of the phase quadrature of the light field $\mathrm{Var}(\hat{Y}_1)$, since $\mathrm{Var}(\hat{Y}_1) = \mathrm{Var}(\hat{Y}_2)$ and $\mathrm{Cov}(\hat{Y}_2,\hat{Y}_1) = 0$, thus $\mathrm{Var}(\hat{S}_b) = 2\mathrm{Var}(\hat{Y}_1)$. At the end we calculate the squeezing parameter, which in our case ($\theta = 0$) is given by:

\begin{align}
\xi_{s_2} = \sqrt{N_a}\frac{\sqrt{\mathrm{Var}(\hat{S}_2)}}{\langle \hat{J}_x\rangle}
\label{xi_s_def}
\end{align}

\section{No Spontaneous Emission}

\subsection{Atomic expectation values}\label{atom_appendix_ns}

The atomic equations with no spontaneous emission are given by:

\begin{align}
\hat{a}_1(t) = \hat{a}_1(0)e^{i\frac{g^2}{\Delta}\int_{0}^{t}\hat{b}_{01}^\dagger(t')\hat{b}_{01}(t')dt'}
\end{align}

\begin{align}
\hat{a}_1^\dagger(t) = \hat{a}_1^\dagger(0)e^{-i\frac{g^2}{\Delta}\int_{0}^{t}\hat{b}_{01}^\dagger(t')\hat{b}_{01}(t')dt'}
\end{align}

Hence, the atomic population operator is independent of time:

\begin{align}
\hat{N}_{a_1}(t) = \hat{a}_1^\dagger(t)\hat{a}_1(t) = \hat{a}_1^\dagger(0)\hat{a}_1(0)
\end{align}

We consider that our total state initially is given by the product:
\begin{align}
|\Psi\rangle = |\alpha_1\rangle\otimes|\alpha_2\rangle\otimes|\beta_1\rangle\otimes|\beta_2\rangle\otimes|0\rangle
\end{align}

meaning that the atomic ensemble as well as the two light fields are in coherent states while the bath is described by the vacuum state, giving the following expectation values:

\begin{align}
\hat{a}_1(0)|\alpha_1\rangle = \sqrt{\frac{N_a}{2}}|\alpha_1\rangle,
\qquad
\hat{b}_{01}(t)|\beta_1\rangle = \beta_0|\beta_1\rangle,
\qquad
\hat{q}_{1_{\mathrm{in}}}(t)|0\rangle = 0|0\rangle \nonumber \\[5pt]
\hat{a}_2(0)|\alpha_2\rangle = \sqrt{\frac{N_a}{2}}|\alpha_2\rangle
\qquad
\hat{b}_{02}(t)|\beta_2\rangle = \beta_0|\beta_2\rangle,
\qquad
\hat{q}_{2_{in}}(t)|0\rangle = 0|0\rangle
\end{align}

where we have used again $\hat{b}_{0j}(t) = \hat{b}_{j}(z_L,t)$ with $j = 1,2$ for simplicity, and we have considered that $\alpha_1(0) = \alpha_2(0) = \sqrt{N_a/2}$ and $\hat{b}_{01}(t) = \hat{b}_{02}(t) = \beta_0$. Now it is really simple to calculate the atomic expectation values in that case:

\begin{align}
\langle\hat{N}_{a_1}(t)\rangle = \frac{N_a}{2},
\qquad
\langle\hat{N}_{a_1}^2(t)\rangle = \langle\hat{N}_{a_1}^2(t')\rangle =  \langle\hat{N}_{a_1}^2(0)\rangle = \frac{N_a}{2} + \frac{N_a^2}{4}
\label{atom_expectation_nse}
\end{align}

\subsection{Phase Quadrature}

The light equation in the case of no spontaneous emission is:

\begin{align}
\hat{b}_1(z_R,t) = \hat{b}_{01}(t)e^{i\frac{g^2}{c\Delta}\hat{a}_1^\dagger(t)\hat{a}_1(t)}
\end{align}

We select a specific mode of the light field:

\begin{align}
\hat{\mathcal{b}}_1(\tau) = \frac{\sqrt{c}}{\sqrt{\tau}}\int_{0}^{\tau}\hat{b}_1(z_R,t)dt
\end{align}

Here the atomic population is constant, thus:

\begin{align}
\hat{\mathcal{b}}_1(\tau) = \frac{\sqrt{c}}{\sqrt{\tau}}e^{i\frac{g^2}{c\Delta}\hat{a}_1^\dagger(t)\hat{a}_1(t)}\int_{0}^{\tau}\hat{b}_{01}(t)dt
\end{align}

We know that the incident light field obeys the following commutation relation $[\hat{b}_{01}(t),\hat{b}_{01}^\dagger(t')] = \frac{1}{c}\delta(t - t')$ . We find the phase quadrature of the specific mode:

\begin{align}
\hat{Y}_1(\tau) \equiv i(\hat{\mathcal{b}}_1(\tau) - \hat{\mathcal{b}}_1^\dagger(\tau)) = i\frac{\sqrt{c}}{\sqrt{\tau}}\left(e^{i\frac{g^2}{c\Delta}\hat{a}_1^\dagger(t)\hat{a}_1(t)}\int_{0}^{\tau}\hat{b}_{01}(t)dt - e^{-i\frac{g^2}{c\Delta}\hat{a}_1^\dagger(t)\hat{a}_1(t)}\int_{0}^{\tau}\hat{b}_{01}^\dagger(t)dt\right)
\end{align}

We make the small angle approximation: 

\begin{align}
\frac{g^2}{c\Delta}\hat{a}_1^\dagger(t)\hat{a}_1(t) << 1
\end{align}

and we get

\begin{align}
\hat{Y}_1(\tau) \approx \hat{Y}_{1_{\mathrm{in}}}(\tau) - \frac{g^2}{\sqrt{c\tau}\Delta}\hat{a}_1^\dagger(\tau)\hat{a}_1(\tau)\int_{0}^{\tau}\left(\hat{b}_{01}(t) + \hat{b}_{01}^\dagger(t)\right)dt
\label{y1_op_ns}
\end{align}

where 

\begin{align}
\hat{Y}_{1_{\mathrm{in}}}(\tau) = i\frac{\sqrt{c}}{\sqrt{\tau}}\int_{0}^{\tau}\left(\hat{b}_{01}(t) - \hat{b}_{01}^\dagger(t)\right)dt
\end{align}

We calculate the expectation value of the phase quadrature:

\begin{align}
\langle\hat{Y}_1(\tau)\rangle \approx -\frac{g^2N_a\beta_0\tau}{\Delta\sqrt{c\tau}}
\label{Y1_mean_ns}
\end{align}

where we have used that $\langle\hat{Y}_{1_{\mathrm{in}}}(\tau)\rangle = 0$ and assumed that $\beta_0 = \beta_0^*$. We calculate the square of the phase quadrature:

\begin{align}
\hat{Y_1}^2(\tau) \approx \hat{Y}_{1_{\mathrm{in}}}^2(\tau) + \frac{g^4}{c\tau\Delta^2}\hat{a}_1^\dagger(\tau)\hat{a}_1(\tau)\hat{a}_1^\dagger(\tau)\hat{a}_1(\tau)\int_{0}^{\tau}\int_{0}^{\tau}dtdt'\; \left(\hat{b}_{01}(t) + \hat{b}_{01}^\dagger(t)\right)\left(\hat{b}_{01}(t') + \hat{b}_{01}^\dagger(t')\right)
\end{align}

For simplicity we calculate separately

\begin{align}
Q_1 = \int_{0}^{\tau}\int_{0}^{\tau}dtdt'\; \left(\hat{b}_{01}(t)\hat{b}_{01}(t') + \hat{b}_{01}(t)\hat{b}_{01}^\dagger(t') + \hat{b}_{01}^\dagger(t)\hat{b}_{01}(t') + \hat{b}_{01}^\dagger(t)\hat{b}_{01}^\dagger(t')\right)
\end{align}

After using the commutation relation $\left[\hat{b}_{01}(t),\hat{b}_{01}^\dagger(t')\right] = \frac{1}{c}\delta(t-t')$
and the delta function property $\int_{0}^{\tau}\delta(t-t')dt' = 1$ we obtain:

\begin{align}
\langle Q_1\rangle = 4\beta_0^2\tau^2 + \frac{\tau}{c}
\end{align}

Making use of the same commutation relation and the same property of the delta function we find that $\langle\hat{Y}_{1_{\mathrm{in}}}^2(\tau)\rangle = 1$. Thus,

\begin{align}
\langle\hat{Y}_1^2(\tau)\rangle \approx 1 + \frac{g^4}{c\tau\Delta^2}\left(\frac{N_a}{2} + \frac{N_a^2}{4}\right)\left(4\beta_0^2\tau^2 + \frac{\tau}{2c}\right)
\label{Y1_sq_mean_ns}
\end{align}

where we have used Eq.~(\ref{atom_expectation_nse}). For simplicity we can ignore the last term of Eq.~(\ref{Y1_sq_mean_ns}) since $4\beta_0^2\tau^2 >> \tau/2c$:

\begin{align}
\langle\hat{Y}_1^2(\tau)\rangle \approx 1 + \frac{4g^4\beta_0^2\tau^2}{c\tau\Delta^2}\left(\frac{N_a}{2} + \frac{N_a^2}{4}\right)
\label{Y1_sq_mean_ns_approx}
\end{align}

From Eq.~(\ref{Y1_mean_ns}) we have:

\begin{align}
\langle\hat{Y}_1(\tau)\rangle^2 \approx \frac{g^4\beta_0^2N_a^2\tau^2}{c\tau\Delta^2}
\end{align}

Hence, we finally have:

\begin{align}
\mathrm{Var}(\hat{Y}_1(\tau)) \approx 1 + 2\chi_{\mathrm{ns}}^2N_aN_{\mathrm{ph}}
\end{align}

and 

\begin{align}
\mathrm{Var}(\hat{S}_b) = 2\mathrm{Var}(\hat{Y}_1(\tau)) \approx 2 + 4\chi_{\mathrm{ns}}^2N_aN_{\mathrm{ph}}
\label{var_sb_ns}
\end{align}

where we have defined

\begin{align}
\chi_{\mathrm{ns}} \equiv \frac{g^2}{c\Delta},
\qquad
N_{\mathrm{ph}} \equiv \beta_0^2\tau
\end{align}

\subsection{Covariances}

The covariance of $\hat{J}_z(\tau)$ and $\hat{S}_b(\tau)$ is defined as:

\begin{align}
\mathrm{Cov}(\hat{J}_z(\tau),\hat{S}_b(\tau)) = \langle\hat{J}_z(\tau)\hat{S}_b(\tau)\rangle - \langle\hat{J}_z(\tau)\rangle\langle\hat{S}_b(\tau)\rangle
\end{align}

We know that $\langle\hat{S}_b(\tau)\rangle = 0$ , since $\hat{S}_b = \hat{Y}_2 - \hat{Y}_1$. Hence:

\begin{align}
\mathrm{Cov}(\hat{J}_z(\tau),\hat{S}_b(\tau)) = \langle\hat{J}_z(\tau)\hat{Y}_2(\tau)\rangle - \langle\hat{J}_z(\tau)\hat{Y}_1(\tau)\rangle
\end{align}

Using $\hat{J}_z(\tau) = (\hat{N}_{a_1}(\tau) -\hat{N}_{a_2}(\tau))/2$, Eq.~(\ref{y1_op_ns}) and the atomic expectation values from Sec.~(\ref{atom_appendix_ns}) we obtain:

\begin{align}
\mathrm{Cov}(\hat{J}_z(\tau),\hat{S}_b(\tau)) \approx \chi_{\mathrm{ns}}N_a\sqrt{N_{\mathrm{ph}}}
\label{cov_ns}
\end{align}

\subsection{Quantum-enhancement parameter $\xi_s$}

Inserting Eq.~(\ref{cov_ns}) and (\ref{var_sb_ns}) in (\ref{Var_S_2}) we obtain:

\begin{align}
\mathrm{Var}(\hat{S}_2(\tau)) \approx \frac{N_a}{4}\left(1 - \frac{\chi_{\mathrm{ns}}^2N_{\mathrm{ph}}N_a}{\chi_{\mathrm{ns}}^2N_{\mathrm{ph}}N_a + 1/2}\right)
\end{align}

Using the atomic equations of motion we find:

\begin{align}
\langle\hat{J}_x(\tau)\rangle = \frac{N_a}{2}e^{-\chi_{\mathrm{ns}}^2N_{\mathrm{ph}}}
\end{align}

Finally, from Eq.~(\ref{xi_s_def}) we obtain: 

\begin{align}
\xi_{s_2}^{ns}(\tau) \approx e^{\chi_{\mathrm{ns}}^2N_{\mathrm{ph}}}\left(1 - \frac{\chi_{\mathrm{ns}}^2N_{\mathrm{ph}}N_a}{\chi_{\mathrm{ns}}^2N_{\mathrm{ph}}N_a + 1/2}\right)^{1/2}
\end{align}

\section{Spontaneous Emission}

\subsection{Atomic expectation values}

In the case where we have incorporated spontaneous emission the calculation of the atomic expectation values is more complicated, since we use the following atomic equations:

\begin{align}
\hat{a}_1(t) &= \hat{a}_1(0)e^{ig^2(\Omega + i\Gamma)\int_{0}^{t}\hat{b}_{01}^\dagger(t')\hat{b}_{01}(t')dt'} + \nonumber \\[5pt]
&+ \frac{g\sqrt{\gamma_3}}{\Delta-i\gamma_3/2}e^{ig^2(\Omega + i\Gamma)\int_{0}^{t}\hat{b}_{01}^\dagger(t')\hat{b}_{01}(t')dt'}\int_{0}^{t}dt'\hat{b}_{01}^\dagger(t')\hat{q}_{1_{\mathrm{in}}}(t')e^{-ig^2(\Omega + i\Gamma)\int_{0}^{t'}\hat{b}_{01}^\dagger(t'')\hat{b}_{01}(t'')dt''}
\label{a1_op}
\end{align}

\begin{align}
\hat{a}_1^\dagger(t) &= \hat{a}_1^\dagger(0)e^{-ig^2(\Omega - i\Gamma)\int_{0}^{t}\hat{b}_{01}^\dagger(t')\hat{b}_{01}(t')dt'} + \nonumber \\[5pt]
&+\frac{g\sqrt{\gamma_3}}{\Delta + i\gamma_3/2}e^{-ig^2(\Omega - i\Gamma)\int_{0}^{t}\hat{b}_{01}^\dagger(t')\hat{b}_{01}(t')dt'}\int_{0}^{t}dt'\hat{b}_{01}(t')\hat{q}_{1_{\mathrm{in}}}^\dagger(t')e^{ig^2(\Omega - i\Gamma)\int_{0}^{t'}\hat{b}_{01}^\dagger(t'')\hat{b}_{01}(t'')dt''}
\label{a1_dag_op}
\end{align}

For simplicity we assume that the intensity operator in the exponentials does not depend on time, namely is a constant number $\hat{b}_{01}^\dagger(t)\hat{b}_{01}(t) \approx \beta_0^2$. We essentially assume here that the atomic loss is due to the average field intensity.   We also ignore the unitary part of the exponentials, since they would cancel out during the calculation of the atomic expectation values. So, we finally have:

\begin{align}
\hat{a}_1(t) = \underbrace{\sqrt{\epsilon(t)}\hat{a}_1(0)}_{\hat{A}_1(t)} + \underbrace{\frac{g\sqrt{\gamma_3}}{\Delta - i\gamma_3/2}\sqrt{\epsilon(t)}\int_{0}^{t}\sqrt{\epsilon^{-1}(t')}\hat{b}_{01}^\dagger(t')\hat{q}_{1_{\mathrm{in}}}(t')dt'}_{\hat{A}_2(t)}
\label{a1_op_approx}
\end{align}

\begin{align}
\hat{a}_1(t) = \underbrace{\sqrt{\epsilon(t)}\hat{a}_1^\dagger(0)}_{\hat{A}_1^\dagger(t)} + \underbrace{\frac{g\sqrt{\gamma_3}}{\Delta + i\gamma_3/2}\sqrt{\epsilon(t)}\int_{0}^{t}\sqrt{\epsilon^{-1}(t')}\hat{b}_{01}(t')\hat{q}_{1_{\mathrm{in}}}^\dagger(t')dt'}_{\hat{A}_2^\dagger(t)}
\label{a1_dag_op_approx}
\end{align}

where we have defined 

\begin{align}
\epsilon(t) \equiv e^{-2g^2\Gamma \beta_0^2 t}
\end{align}

We calculate the expectation value of atoms in state $|1\rangle$:

\begin{align}
\langle\hat{N}_{a_1}(t)\rangle = \langle\hat{a}_1^\dagger(t)\hat{a}_1(t)\rangle = \frac{N_a}{2}\epsilon(t)
\end{align}

where $\epsilon(t)$ indicates the atomic rate of loss in our system at time $t$. Now we are going to calculate the more complicated expectation value $\langle \hat{N}_{a_1}(t)\hat{N}_{a_1}(t')\rangle$. We have named each term of Eq.~(\ref{a1_op_approx}) and (\ref{a1_dag_op_approx}) for simplicity, in order to clearly show which terms finally survive:

\begin{align}
\langle \hat{N}_{a_1}(t)\hat{N}_{a_1}(t')\rangle = \langle\hat{A}_1^\dagger(t)\hat{A}_1(t)\hat{A}_1^\dagger(t')\hat{A}_1(t')\rangle + \langle\hat{A}_1^\dagger(t)\hat{A}_2(t)\hat{A}_2^\dagger(t')\hat{A}_1(t')\rangle
\label{Na1t_Na1tdash}
\end{align}

where all the other terms in this product are zero since $\langle\hat{q}_{1_{\mathrm{in}}}\rangle =  \langle\hat{q}_{1_{\mathrm{in}}}^\dagger\rangle = \langle\hat{q}_{1_{\mathrm{in}}}^\dagger\hat{q}_{1_{\mathrm{in}}}\rangle = 0$. The first term of the above equation is easily calculated:

\begin{align}
\langle\hat{A}_1^\dagger(t)\hat{A}_1(t)\hat{A}_1^\dagger(t')\hat{A}_1(t')\rangle = \left(\frac{N_a}{2} + \frac{N_a^2}{4}\right)\epsilon(t)\epsilon(t')
\label{simple_term}
\end{align}

However the second term is more complicated:

\begin{align}
\langle\hat{A}_1^\dagger(t)\hat{A}_2(t)\hat{A}_2^\dagger(t')\hat{A}_1(t')\rangle &=\; 2g^2\Gamma\epsilon(t)\epsilon(t')\langle\hat{a}_1^\dagger(0)\hat{a}_1(0)\rangle\times\nonumber \\[5pt] &\times\int_{0}^{t}\int_{0}^{t'}d\xi ds \; \sqrt{\epsilon^{-1}(s)}\sqrt{\epsilon^{-1}(\xi)}\langle\hat{b}_{01}^\dagger(s)\hat{b}_{01}(\xi)\rangle\langle\hat{q}_{1_{\mathrm{in}}}(s)\hat{q}_{1_{\mathrm{in}}}^\dagger(\xi)\rangle
\end{align}

We use the commutation relation for the temporal part of the Langevin noise:

\begin{align}
\left[\hat{q}_{1_{\mathrm{in}}}(s),\hat{q}_{1_{\mathrm{in}}}^\dagger(\xi)\right] = \delta(\xi - s)
\end{align}

We also make use of the following property of the delta function:

\begin{align}
\int_{0}^{t'}d\xi f(\xi)\delta(\xi - s) = f(s)\Theta(t'-s)
\end{align}

where $\Theta(t'-s)$ is the Heaviside step function and using $\langle\hat{b}_{01}^\dagger(s)\hat{b}_{01}(s)\rangle = \beta_0^2$ we obtain

\begin{align}
\langle\hat{A}_1^\dagger(t)\hat{A}_2(t)\hat{A}_2^\dagger(t')\hat{A}_1(t')\rangle =\; g^2\Gamma N_a\beta_0^2\epsilon(t)\epsilon(t')\int_{0}^{t}ds\; \epsilon^{-1}(s)\Theta(t'-s)
\label{complicated_term}
\end{align}

For $t\ge t'$ we have:

\begin{align}
\langle\hat{A}_1^\dagger(t)\hat{A}_2(t)\hat{A}_2^\dagger(t')\hat{A}_1(t')\rangle =\; \frac{N_a}{2}\epsilon(t)\left(1 - \epsilon(t')\right)
\label{complicated_term2}
\end{align}

and using Eq.~(\ref{Na1t_Na1tdash}), (\ref{simple_term}) and (\ref{complicated_term2}) we get:

\begin{align}
\langle \hat{N}_{a_1}(t)\hat{N}_{a_1}(t')\rangle = \frac{N_a^2}{4}\epsilon(t)\epsilon(t') + \frac{N_a}{2}\epsilon(t)
\label{Na1_t_t'_mean1}
\end{align}

While for $t < t'$ we have:

\begin{align}
\langle\hat{A}_1^\dagger(t)\hat{A}_2(t)\hat{A}_2^\dagger(t')\hat{A}_1(t')\rangle =\; \frac{N_a}{2}\epsilon(t')\left(1 - \epsilon(t)\right)
\end{align}

and

\begin{align}
\langle \hat{N}_{a_1}(t)\hat{N}_{a_1}(t')\rangle = \frac{N_a^2}{4}\epsilon(t)\epsilon(t') + \frac{N_a}{2}\epsilon(t')
\label{Na1_t_t'_mean2}
\end{align}

We notice that we obtain the same result for the double integral with respect to $t$ and $t'$ for both cases, $t\ge t'$ and for $t<t'$

\begin{align}
\int_{0}^{\tau}\int_{0}^{\tau}dt\;dt'\; \langle \hat{N}_{a_1}(t)\hat{N}_{a_1}(t')\rangle = \frac{N_a^2}{4}I_1^2 + \frac{N_a}{2}I_1\tau
\label{int_Na1t_Na1tdash}
\end{align}

but distinguishing between the two cases would be important when we calculate the covariance of $\hat{J}_z(\tau)$ and $\hat{S}_b(\tau)$. For simplicity we have also defined:

\begin{align}
I_1(\tau) \equiv \int_{0}^{\tau}dt\;\epsilon(t) = \frac{1 - \epsilon(\tau)}{2g^2\Gamma\beta_0^2}
\end{align}

We also calculate:

\begin{align}
\int_{0}^{\tau} dt \langle\hat{N}_{a_1}(t)\rangle = \frac{N_a}{2}I_1 
\label{int_Na1_mean}
\end{align}

\subsection{Phase Quadrature}

In the case of spontaneous emission the photon operator is given by the following equation:

\begin{align}
\hat{b}_1(z_R,t) = \hat{b}_{01}(t)e^{i\frac{g^2}{c}(\Omega + i\Gamma)\hat{a}_1^\dagger(t)\hat{a}_1(t)} + \frac{g}{c}\frac{\sqrt{\gamma_3}}{\Delta - i\gamma_3/2}\hat{a}_1^\dagger(t)\hat{q}_{1_{\mathrm{in}}}(t)
\label{light_eq}
\end{align}

Again we define the phase quadrature operator of a specific mode of the light field:

\begin{align}
\hat{Y}_1(\tau) = i(\hat{\mathcal{b}}_1(\tau) - \hat{\mathcal{b}}_1^\dagger(\tau))
\end{align}

where

\begin{align}
\hat{\mathcal{b}}_1(t) = \frac{\sqrt{c}}{\sqrt{\tau}}\int_{0}^{\tau}\hat{b}_{01}(t)dt
\end{align}

Making the small angle approximation $g^2\left(\Omega + i\Gamma\right)\hat{a}_1^\dagger\hat{a}_1/c << 1$ we obtain:

\begin{align}
\hat{Y}_1 \approx \hat{Y}_{1_{\mathrm{in}}} - \frac{g^2\Omega}{\sqrt{c\tau}}\int_{0}^{\tau}\left(\hat{b}_{01}(t) + \hat{b}_{01}^\dagger(t)\right)\hat{a}_1^\dagger(t)\hat{a}_1(t)dt  - \frac{g^2\Gamma}{\sqrt{c\tau}}\int_{0}^{\tau}\left(\hat{b}_{01}(t) - \hat{b}_{01}^\dagger(t)\right)\hat{a}_1^\dagger(t)\hat{a}_1(t)dt \; + \nonumber \\[5pt] + \; i\frac{g\Omega\sqrt{\gamma_3}}{\sqrt{c\tau}}\int_{0}^{\tau}dt\left(\hat{q}_{1_{\mathrm{in}}}(t)\hat{a}_1^\dagger(t) - \hat{q}_{1_{\mathrm{in}}}^\dagger(t)\hat{a}_1(t)\right) - i\frac{g\Gamma\sqrt{\gamma_3}}{\sqrt{c\tau}}\int_{0}^{\tau}dt\left(\hat{q}_{1_{\mathrm{in}}}(t)\hat{a}_1^\dagger(t) + \hat{q}_{1_{\mathrm{in}}}^\dagger(t)\hat{a}_1(t)\right) 
\label{y1_op_approx}
\end{align}

where 

\begin{align}
\hat{Y}_{1_{\mathrm{in}}}(\tau) \equiv i\frac{\sqrt{c}}{\sqrt{\tau}}\int_{0}^{\tau}dt\left(\hat{b}_{01}(t) - \hat{b}_{01}^\dagger(t)\right)
\end{align}

We calculate the expectation value of $\hat{Y}_1(\tau)$:

\begin{align}
\langle\hat{Y}_1(\tau)\rangle \approx -\frac{2\beta_0\Omega g^2}{\sqrt{c\tau}}\int_{0}^{\tau}dt\langle\hat{N}_{a_1}(t)\rangle
\label{Y1_mean}
\end{align}

From Eq.~(\ref{int_Na1_mean}) we get:

\begin{align}
\langle\hat{Y}_1\rangle^2 \approx \frac{g^4\Omega^2\beta_0^2N_a^2I_1^2}{c\tau}
\label{phase_quad_sq_se}
\end{align}

Now we are going to calculate $\langle \hat{Y}_1^2\rangle$, where for simplicity we keep only the terms coming from the the first two terms of Eq.~(\ref{y1_op_approx}), since they are the dominant terms :

\begin{align}
\langle \hat{Y}_1^2\rangle \approx 1 + \frac{4g^4\Omega^2}{c\tau}\beta_0^2\int_{0}^{\tau}\int_{0}^{\tau}dtdt'\langle\hat{N}_{a_1}(t')\hat{N}_{a_1}(t)\rangle
\label{Y_sq}
\end{align}

Substituting Eq.~(\ref{int_Na1t_Na1tdash}) in (\ref{Y_sq}) and using (\ref{phase_quad_sq_se}) we obtain:

\begin{align}
\mathrm{Var}(\hat{Y}_1(\tau)) \approx 1 + 2\chi_1^2N_{\mathrm{ph}}N_a\overline{\epsilon(\tau)}
\label{Var_Y1}
\end{align}

where we have defined $\chi_1\equiv \frac{g^2\Omega}{c}$ and $\overline{\epsilon(\tau)} = \frac{1}{\tau}\int_{0}^{\tau}\epsilon(t)dt$ which is the time average of the decay. We notice that $\chi_1 = \chi_{\mathrm{ns}}$ in the no spontaneous emission case ($\gamma_3=0$). As we mentioned before $\mathrm{Var}(\hat{S}_b) = 2\mathrm{Var}(\hat{Y}_1)$, thus:

\begin{align}
\mathrm{Var}(\hat{S}_b(\tau)) \approx 2 + 4\chi_1^2N_{\mathrm{ph}}N_a\overline{\epsilon(\tau)}
\label{Var_Sb}
\end{align}

\subsection{Covariances}

The covariance of $\hat{J}_z$ and $\hat{S}_b$ is again given by $\mathrm{Cov}(\hat{J}_z(\tau),\hat{S}_b(\tau)) = \langle\hat{J}_z(\tau)\hat{Y}_2(\tau)\rangle - \langle\hat{J}_z(\tau)\hat{Y}_1(\tau)\rangle$, which gives

\begin{align}
\mathrm{Cov}(\hat{J}_z(\tau),\hat{S}_b(\tau)) = \frac{2g^2\Omega\beta_0}{\sqrt{c\tau}}\int_{0}^{\tau}dt\left(\langle\hat{N}_{a_1}(\tau)\hat{N}_{a_1}(t)\rangle - \langle\hat{N}_{a_1}(\tau)\hat{N}_{a_2}(t)\rangle \right)
\label{Cov_Jz_Sb1}
\end{align}

since

\begin{align}
\langle\hat{N}_{a_1}(\tau)\hat{N}_{a_1}(t)\rangle  = \langle\hat{N}_{a_2}(\tau)\hat{N}_{a_2}(t)\rangle
\qquad
\langle\hat{N}_{a_1}(\tau)\hat{N}_{a_2}(t)\rangle = \langle\hat{N}_{a_2}(\tau)\hat{N}_{a_1}(t)\rangle
\end{align}

Now we have to be a bit more careful, compared to the no spontaneous emission case, because we have two different expressions for $\langle \hat{N}_{a_1}(t)\hat{N}_{a_1}(t')\rangle$ depending on whether $t\ge t'$ or $t<t'$. That's why we are going to calculate $\mathrm{Cov}(\hat{S}_b(\tau),\hat{J}_z(\tau))$ as well:

\begin{align}
\mathrm{Cov}(\hat{S}_b(\tau),\hat{J}_z(\tau)) = \frac{2g^2\Omega\beta_0}{\sqrt{c\tau}}\int_{0}^{\tau}dt\left(\langle\hat{N}_{a_1}(t)\hat{N}_{a_1}(\tau)\rangle - \langle\hat{N}_{a_1}(t)\hat{N}_{a_2}(\tau)\rangle \right)
\end{align}

For the first covariance, where $\tau\ge t$ we use Eq.~(\ref{Na1_t_t'_mean1}), hence:

\begin{align}
\int_{0}^{\tau}dt\;\langle\hat{N}_{a_1}(\tau)\hat{N}_{a_1}(t)\rangle = \frac{N_a^2}{4}\epsilon(\tau)I_1 + \frac{N_a}{2}\epsilon(\tau)\tau
\end{align}

We calculate the simpler term:

\begin{align}
\langle\hat{N}_{a_1}(\tau)\hat{N}_{a_2}(t)\rangle = \frac{N_a^2}{4}\epsilon(\tau)\epsilon(t)
\label{Na1_Na2_mean}
\end{align}

since $\hat{a}_1(t)$ commutes with $\hat{a}_2(t')$ for all $t$ and $t'$. Thus,

\begin{align}
\int_{0}^{\tau}dt\;\langle\hat{N}_{a_1}(\tau)\hat{N}_{a_2}(t)\rangle = \frac{N_a^2}{4}\epsilon(\tau)I_1
\end{align}

We finally have:

\begin{align}
\mathrm{Cov}(\hat{J}_z(\tau),\hat{S}_b(\tau)) = \chi_1\sqrt{N_{\mathrm{ph}}}N_a\epsilon(\tau)
\label{Cov_Jz_Sb}
\end{align}

where we used again $\chi = \frac{g^2\Omega}{c}$. For the second covariance we use Eq.~(\ref{Na1_t_t'_mean2}) for $t<t'$ and we obtain:

\begin{align}
\int_{0}^{\tau}dt\;\langle\hat{N}_{a_1}(t)\hat{N}_{a_1}(\tau)\rangle = \frac{N_a^2}{4}\epsilon(\tau)I_1 + \frac{N_a}{2}\epsilon(\tau)\tau
\end{align}

\begin{align}
\int_{0}^{\tau}dt\;\langle\hat{N}_{a_1}(t)\hat{N}_{a_2}(\tau)\rangle = \frac{N_a^2}{4}\epsilon(\tau)I_1
\end{align}

Hence, we finally get the same result for both covariances as we expected:

\begin{align}
\mathrm{Cov}(\hat{J}_z(\tau),\hat{S}_b(\tau)) = \mathrm{Cov}(\hat{S}_b(\tau),\hat{J}_z(\tau) = \chi_1\sqrt{N_{\mathrm{ph}}}N_a\epsilon(\tau)
\label{both_covariances}
\end{align}

\subsection{Quantum-enhancement parameter $\xi_s$}\label{appendix_final}

Substituting Eq.~(\ref{Var_Sb}) and (\ref{both_covariances}) into Eq.~(\ref{Var_S_2}) we get:

\begin{align}
\mathrm{Var}(\hat{S}_2(\tau)) \approx \frac{N_a}{4}\epsilon(\tau)\left(1 - \frac{\chi_1^2N_{\mathrm{ph}}N_a\epsilon(\tau)}{\chi_1^2N_{\mathrm{ph}}N_a\overline{\epsilon(\tau)} + 1/2}\right)
\end{align} 

Using the atomic equations we find the expectation value of $\hat{J}_x$:

\begin{align}
\langle \hat{J}_x\rangle = \frac{N_a}{2}e^{-(\chi_1^2 + 2\chi_2)N_{\mathrm{ph}}}
\end{align}

where we have defined $\chi_2\equiv g^2\Gamma/c$. Now we can express $\epsilon(\tau)$ in a more convenient way $\epsilon(\tau) = e^{-2\chi_2N_{\mathrm{ph}}}$. Finally the squeezing parameter is given by:

\begin{align}
\xi_{s_2} \approx e^{(\chi_1^2 + \chi_2)N_{\mathrm{ph}}}\left(1 - \frac{\chi_1^2N_{\mathrm{ph}}N_a\epsilon(\tau)}{\chi_1^2N_{\mathrm{ph}}N_a\overline{\epsilon(\tau)} + 1/2}\right)^{1/2}
\label{ksi_s2_se}
\end{align} 

where for convenience we present again all the parameter definitions we made throughout this calculation:

\begin{align}
\chi_1 \equiv \frac{g^2\Omega}{c},
\qquad
\chi_2 \equiv \frac{g^2\Gamma}{c},
\qquad
\overline{\epsilon(\tau)} \equiv \frac{1}{\tau}\int_{0}^{\tau}\epsilon(t)dt,
\qquad
\epsilon(\tau) = e^{-2\chi_2N_{\mathrm{ph}}}
\end{align}

\end{document}